\documentclass{IEEEtran}
\usepackage{amsmath,amsfonts,amsthm,amssymb}
\usepackage{algorithmic}
\usepackage{array}
\usepackage{textcomp}
\usepackage{stfloats}
\usepackage{enumerate}
\usepackage{bm,bbm}
\usepackage{url}
\usepackage{verbatim}
\usepackage{graphicx}
\usepackage[noadjust]{cite}
\usepackage{subcaption}
\usepackage{multirow}
\usepackage{balance}

\usepackage{autobreak}
\theoremstyle{plain}
\newtheorem{theorem}{Theorem}[section]
\newtheorem{proposition}[theorem]{Proposition}
\newtheorem{lemma}[theorem]{Lemma}

\theoremstyle{definition}

\theoremstyle{remark}
\newtheorem{remark}[theorem]{Remark}

\newcommand\norm[1]{\lVert#1\rVert}

\begin{document}
\title{Stable estimation of pulses of unknown shape from multiple snapshots via ESPRIT}
\author{Meghna Kalra and Kiryung Lee,~\IEEEmembership{Senior Member,~IEEE}\thanks{This work was supported in part by NSF CAREER Award CCF-1943201. Meghna Kalra and Kiryung Lee are with the Department of Electrical and Computer Engineering at the Ohio State University, Columbus, OH 43220 (e-mail: kalra.42@osu.edu, kiryung@ece.osu.edu). A preliminary version of this work has been presented at the 14th International Conference on Sampling Theory and Applications
(SampTA) \cite{kalra2023stability}.}}

\IEEEoverridecommandlockouts

\maketitle

\begin{abstract}
We consider the problem of resolving overlapping pulses from noisy multi-snapshot measurements, which has been a problem central to various applications including medical imaging and array signal processing. 
ESPRIT algorithm has been used to estimate the pulse locations. 
However, existing theoretical analysis is restricted to ideal assumptions on signal and measurement models. 
We present a novel perturbation analysis that overcomes the previous theoretical limitation, which is derived without a stringent assumption on the signal model.  
Our unifying analysis applies to various sub-array designs of the ESPRIT algorithm. We demonstrate the usefulness of the perturbation analysis by specifying the result in two practical scenarios. 
In the first scenario, we quantify how the number of snapshots for stable recovery scales when the number of Fourier measurements per snapshot is sufficiently large. 
In the second scenario, we propose compressive blind array calibration by ESPRIT with random sub-arrays and provide the corresponding non-asymptotic error bound. 
Furthermore, we demonstrate that the empirical performance of ESPRIT corroborates the theoretical analysis through extensive numerical results. 

\end{abstract}


\section{Introduction}
\IEEEPARstart{W}{e} address the problem of estimating a pulse stream with an unknown shape from noisy Fourier measurements across multiple snapshots with varying amplitudes. 
The pulse locations are shared over all snapshots so that the observed signal at the $l$th snapshot, denoted by $y_l(t)$, is a superposition of $S$ pulses of unknown shape $g(t)$ centered at common locations in $\{\tau_k\}_{k=1}^S$, i.e. 
\begin{equation}
\label{eq:signal_model_td}
    y_l(t) = \sum_{k=1}^S x_{k,l}  g(t-\tau_k), \quad l = 1,\dots,L,
\end{equation}
where $L$ denotes the total number of snapshots. 
We assume that all pulses are supported within the interval $[0,T)$ for some $T > 0$.
The goal is to obtain the locations $\{ \tau_k\}_{k=1}^S$ of the pulses from the noisy Fourier measurements of $y_l(t)$ across $L$ snapshots. 
It is of interest to resolve the pulses of an unknown shape in the presence of noise when they are closely located with significant overlaps. 
This problem arises in various fields such as wireless communications, seismology, sonar imaging, and medical imaging.

The estimation problem has been studied as the resolution of overlapping echos and blind array calibration \cite{paulraj1985direction,wylie1993self,eldar2020sensor,bresler1989resolution,asztely1997auto,astely1999spatial,swindlehurst1999methods}. 
Various estimators have been developed in the following two categories. 
In one approach, based on the assumption that the amplitudes correspond to uncorrelated random variables, algebraic methods leveraging a special structure in the data covariance matrix have been developed \cite{paulraj1985direction,wylie1993self}. 
A recent work provided identifiability conditions and stability analysis of these algebraic methods \cite{eldar2020sensor}. 
However, the uncorrelatedness of magnitudes is often violated in imaging applications.  
In another approach, subspace methods have been developed under a weaker condition that the amplitude matrix has the full row rank \cite{bresler1989resolution,asztely1997auto,swindlehurst1999methods,astely1999spatial}. 
These subspace methods are variants of the Estimation of Signal Parameters via Rotational Invariance Techniques (ESPRIT) algorithm \cite{roy1989esprit}. 
This paper focuses on the scenario with a weaker assumption on the amplitudes considered in the latter approach.

Previously it has been shown that ESPRIT can exactly recovers the locations of overlapping echoes in an ideal case \cite{bresler1989resolution}. 
ESPRIT works on the principle of dividing the data into two sub-arrays related via pulse locations. 
They have shown the success of ESPRIT under the \textit{pairwise identical condition} (PIC), which refers to the pairwise coincidence of the Fourier transform of the pulse shape between the two sub-arrays. 
Therefore, their analysis has been restricted to the satisfaction of the PIC and noise-free measurements. 
Subsequent work proposed subspace methods with improved empirical performances \cite{asztely1997auto,swindlehurst1999methods,astely1999spatial}. 
However, they did not show when their methods will succeed. 
We are motivated to further extend ESPRIT and its existing theoretical analysis \cite{bresler1989resolution} toward practical scenarios. \\\vspace{-5mm}

\noindent\textbf{Contributions: }
We present a perturbation analysis of ESPRIT without reliance on the PIC, derived under the mild assumption that the sub-arrays of ESPRIT are related by a uniform shift. This assumption allows our unifying analysis to apply to all sub-array designs considered in the literature \cite{bresler1989resolution,asztely1997auto,astely1999spatial,swindlehurst1999methods}. The novelty of our work lies in overcoming the limitation of existing theoretical analysis of the subspace methods \cite{bresler1989resolution}. We extend the theory of ESPRIT to the case where the observations are corrupted with noise and the model does not exactly satisfy the PIC. Our work can be considered as an extension of the method presented in \cite{swindlehurst1999methods} by providing a theoretical analysis of ESPRIT without the PIC condition. 
The usefulness of our perturbation analysis is demonstrated by specifying the result in two scenarios arising in practical applications.
The first scenario considers the case where the number of Fourier measurements $M$ can be large, but the number of snapshots $L$ is desired to be small. This case is particularly relevant to applications such as surface electromyography (sEMG), where the cost is associated with increasing $L$, which corresponds to the number of sensors. Our results determine the number of snapshots required for a stable estimate of the location as a function of the model parameters. 
In the second scenario, we explore an application more aligned with blind array calibration. 
In this context, increasing $M$, which corresponds to the number of receivers, carries the cost, whereas $L$ can be arbitrarily large. Specifically, when the pulses are closely located, a larger $M$ is needed to resolve them accurately if the Fourier measurements are taken on a uniform grid \cite{moitra2015super}. 
To address this challenge, we propose a random sub-array design in the vein of the compressed sensing \cite{tang2013compressed} where $M$ samples are randomly chosen at pairs of consecutive indices from a larger aperture pool. 
This random sampling strategy captures more information as compared to the uniform design. 
Furthermore, the rationale for selecting random samples in pairs is to satisfy the rotation invariance structure required by ESPRIT. The novelty of our approach lies in this newly proposed random sub-array design, coupled with the theoretical analysis derives an accompanying error bound. 
Moreover, we provide comprehensive numerical results that corroborate our theoretical findings.\\\vspace{-5mm}

\noindent\textbf{Related Work: }
The classical methods such as Prony's \cite{prony1795essai}, MUSIC (MUltiple SIgnal Classification) \cite{schmidt1986multiple}, and ESPRIT \cite{roy1989esprit} have been used to recover the support of an impulse train from its Fourier measurements. 
Recent papers provided a non-asymptotic analysis of these methods \cite{liao2016music,li2020super,li2021stability}.
Another line of research studied the atomic norm minimization approach in the single snapshot scenario \cite{candes2013super,tang2013compressed,candes2014towards} and in the multi snapshot scenario \cite{li2015off}. 
The recovery of Diracs considered in these papers can be seen as a special instance of our problem formulation where the pulse shape is trivially the Dirac delta distribution. 
These methods are inapplicable when dealing with a pulse shape that is not trivial and unknown.  

\begin{figure} [!ht]
		\begin{subfigure}
      {.45\textwidth}
			\centering
\includegraphics[width=.9\linewidth]{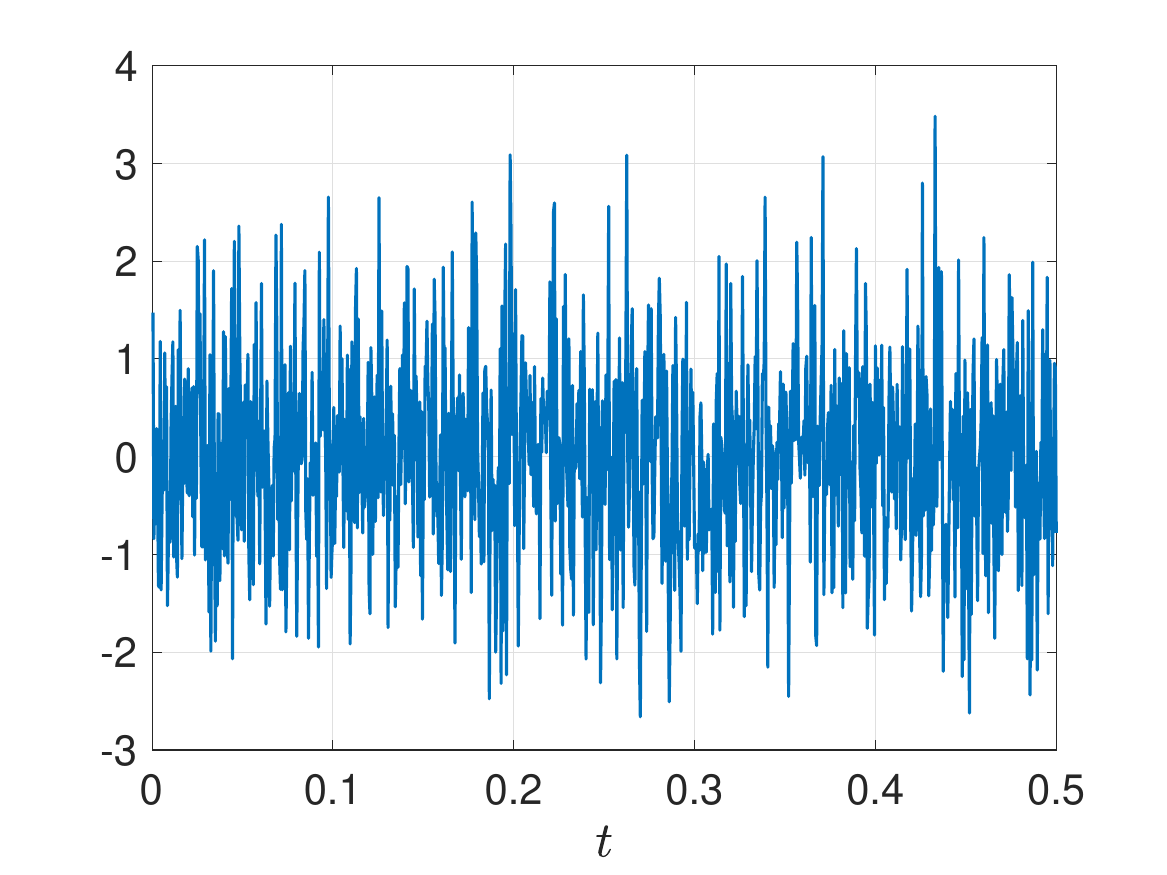}
	\caption{signal from random subspace }
 \label{fig:rand_subspace}
		\end{subfigure}
		\begin{subfigure}{.45\textwidth}
			\centering
\includegraphics[width=.9\linewidth]{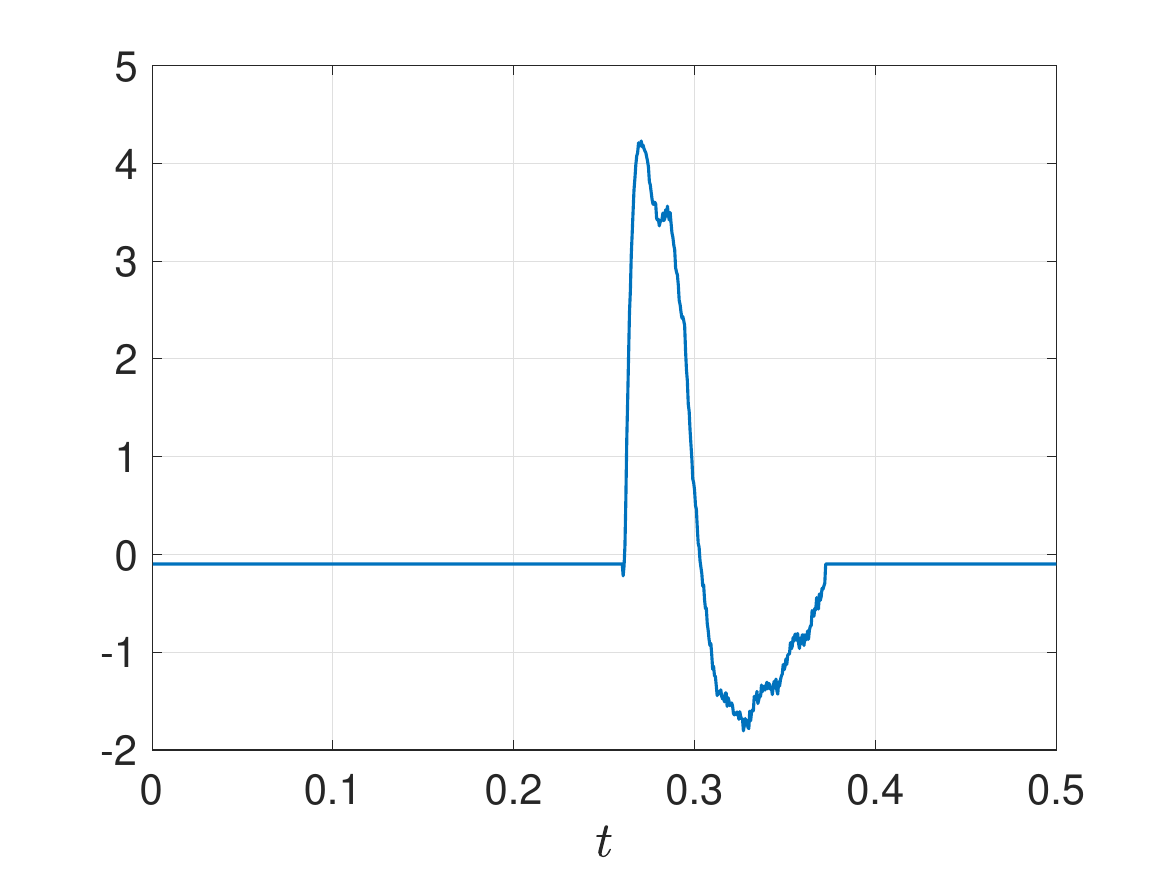}
			\caption{sEMG signal \footnotemark}
   \label{fig:semg signal}
		\end{subfigure}	
  \caption{Illustration of examples of signals from a random subspace and in practical applications.}
\end{figure}

The atomic norm approach has been extended to the blind spike deconvolution \cite{chi2016guaranteed,yang2016super}, which is a variant of the estimation considered in this paper in the single snapshot case under the assumption that the unknown pulse belongs to a known random subspace. 
They established theoretical analysis to ensure performance guarantees for the convex programs.
However, the random subspace assumption is often not suitable for modeling signals in practical applications.  Fig.~\ref{fig:rand_subspace} and Fig.~\ref{fig:semg signal} illustrate an example of signals from a random subspace and sEMG (surface electromyography) signals. 
The sEMG signal can be modeled using \eqref{eq:signal_model_td}, and its appearance is distinct from that of signals from a random subspace, demonstrating that the random subspace assumption is not suitable for signals akin to sEMG. 
In contrast, the theoretical analysis in this paper has been derived from a mild assumption that the pulse shape changes slowly in the Fourier domain. 
A recent paper also studied the problem of blind super-resolution without a stringent assumption on the pulse shape structure \cite{da2019self}, but its theoretical analysis is yet to be explored.

Resolving overlapping echoes is also related to formulations of multi-channel blind deconvolution \cite{wang2016blind,li2018global,shi2021manifold}. 
Theoretical analysis of practical optimization methods has been established when the observations are given as the convolution of an all-pass signal and random sparse filters. 
The estimation problem considered in this paper corresponds to the multi-channel spike deconvolution for sparse continuous-time filters sharing a common support. 
Another related problem considers blind array calibration when the steering matrix is random Gaussian \cite{ling2018self,li2018blind}. 
A different formulation of blind calibration where the observations are given through the modulation with input sequences have been studied \cite{perros2000blind,perros2000pulse}. 
Due to variations in problem formulations, their results are not directly comparable to ours.

\footnotetext{This signal is obtained by bandpass filtering of the measured sEMG signal. The filtering is used to suppress physiological noise (such as breathing and respiration). The location of the pulses in the signal portion is preserved during the filtering. }

\noindent\textbf{Notation: }
The spectral norm of a matrix $\bm{A}$ is denoted as $\norm{\bm{A}}$. The $k$-th eigenvalue or singular value of $\bm{A}$ is denoted $\lambda_k(\bm{A})$ or $\sigma_k(\bm{A})$. The condition number of any matrix $\bm{A}$ is denoted as $\kappa(\bm{A})$. The Moore-penrose inverse of $\bm{A}$ is denoted by $\bm{A}^\dagger$. The maximum of two real numbers $a$ and $b$ will be denoted by $a \vee b$. 
We will use a shorthand notation $[n]$ to denote the set of integers $\{1,\ldots,n\}$. The notation $a\lesssim b$ denotes the relationship between two scalars $a$ and $b$ such that there exists an absolute constant $C>0$ for which $a \leq Cb$. 
The distance between the column spaces by two matrices $\bm{U}$ and $\widehat{\bm{U}}$ is defined by $\mathrm{dist}(\bm{\widehat{U}},{\bm{U}}) :=
\norm{\widehat{\bm U} \widehat{\bm U}^\mathsf{H} - \bm{U} \bm{U}^\mathsf{H}}$, which corresponds to the sine of the largest principal angle.
\\\vspace{-5mm}

\noindent\textbf{Organization: }
The rest of the paper is organized as follows: Section~\ref{sec:det_analysis} presents an overview of ESPRIT for resolving a pulse stream of unknown shape from multiple snapshots, along with the non-asymptotic perturbation analysis. In Section~\ref{sec:max_overlap}, we specify the analysis to a scenario where the cost of snapshots is more crucial than the Fourier measurements and provide an error bound as a function of key model parameters. Section~\ref{sec:random_doublets} illustrates the perturbation analysis on compressive blind array calibration, focusing on cases where the cost of Fourier measurements predominates the cost of obtaining multiple snapshots. The numerical results are presented in Section~\ref{sec:num_res}, and we conclude with a brief summary of contributions and discussions in Section~\ref{sec:conclusion}.

\section{Perturbation analysis of ESPRIT with unknown pulse shape}
\label{sec:det_analysis}
 In this section, we present a non-asymptotic perturbation analysis for the estimation problem of resolving unknown pulses using ESPRIT. We begin with describing the ESPRIT algorithm in the context of our signal model. 
Let $\bm Y \in \mathbb{C}^{|\Omega| \times L}$ collect all noise-free Fourier measurements at frequencies given by $\Omega = \{\omega_i\}_{i=1}^{|\Omega|}$ such that 
\begin{equation}
\label{eq:fourier_meas_model}
\begin{aligned}
(\bm Y)_{i,l} = Y_l(\omega_i) &= \sum_{k=1}^S x_{k,l} G(\omega_i) e^{-\mathsf{j} 2 \pi \tau_k \omega_i}, \\
& \quad\quad\quad\quad i \in [|\Omega|], ~ l \in [L]. 
\end{aligned}
\end{equation}
Then data matrix $\bm{Y}$ in \eqref{eq:fourier_meas_model} can be written compactly as 
\begin{equation}
\label{eq:decomp_noisefree_data}
\bm Y = \bm G \bm \Phi \bm X,     
\end{equation}
where $\bm G \in \mathbb{C}^{|\Omega| \times |\Omega|}$, $\bm \Phi \in \mathbb{C}^{|\Omega| \times S}$, and $\bm X \in \mathbb{C}^{S \times L}$ are defined by
\begin{equation}
\label{eq:def_matrices}
\begin{aligned}
(\bm G)_{i,j} &= G(\omega_i) \delta_{ij}, \quad i,j \in [|\Omega|], \\
(\bm \Phi)_{i,k} &= e^{-\mathsf{j} 2 \pi \tau_k \omega_i}, \quad i \in [|\Omega|], ~ k \in [S], \\
(\bm X)_{k,l} &= x_{k,l}, \quad k \in [S], ~ l \in [L].
\end{aligned}
\end{equation}
The ESPRIT algorithm \cite{roy1989esprit} is built on the special structure of the sampling index set $\Omega$ so that there exist $\Omega_1, \Omega_2 \subset \Omega$ satisfying $\Omega = \Omega_1 \cup \Omega_2$ and $|\Omega_1| = |\Omega_2|$. 
In the context of array signal processing, $\Omega_1$ and $\Omega_2$ are referred to as sub-arrays since they characterize the configuration of the sensor array. 
ESPRIT is a subspace-based method that estimates the pulse locations from an estimate of the column space of $\bm{G}\bm{\Phi}$, which is also known as the signal subspace. 
Suppose that $\bm{X}$ has the full row rank, which implies $L \geq S$. 
Then one can find $\widehat{\bm U} \in \mathbb{C}^{|\Omega| \times S}$ spanning an estimate of the signal subspace from an observed version of $\bm{Y}$ potentially corrupted with noise. 
ESPRIT estimates the pulse locations up to a permutation ambiguity by
\[
\widehat{\tau}_k = - \frac{T \cdot  \arg(\lambda_k(\widehat{\bm U}_1^\dagger \widehat{\bm U}_2))}{2 \pi} 
\]
for $k \in [S]$ with $\widehat{\bm U}_1 = \bm \Pi_1 \widehat{\bm U}$ and $\widehat{\bm U}_2 = \bm \Pi_2 \widehat{\bm U}$, where $\bm \Pi_1, \bm \Pi_2 \in \mathbb{R}^{|\Omega_1| \times |\Omega|}$ are the selector matrices defined by
\[
(\bm \Pi_j)_{m,i} 
= \begin{cases} 
1 & \text{if } \omega_i = \omega_{j,m}, \\ 
0 & \text{else}.
\end{cases}
\]
The eigenvalues of $\widehat{\bm U}_1^\dagger \widehat{\bm U}_2$ denoted by $\{\lambda_k(\widehat{\bm U}_1^\dagger \widehat{\bm U}_2)\}_{k=1}^S$ are not ordered. Once the pulse locations are estimated, the Fourier measurements of the pulse shape in $\bm G$ can be estimated via least squares. 

ESPRIT will find the exact locations of the unknown pulse shape when $\widehat{\bm{U}}$ spans the signal subspace and $\bm{G}\bm{\Phi}$ satisfies the \emph{rotation-invariance} property defined as
\begin{equation}
\label{eq:rotinv_GPhi}
\bm \Pi_2 \bm G \bm \Phi = \bm \Pi_1 \bm G \bm \Phi \bm D.
\end{equation}
Previous work \cite{bresler1989resolution} identified a sufficient condition for \eqref{eq:rotinv_GPhi}, which is summarized as follows: 
Let $\Omega_2 = \{\omega_{2,m}\}_{m=1}^{|\Omega_1|}$ from $\Omega_1 = \{\omega_{1,m}\}_{m=1}^{|\Omega_1|}$ be constructed by a uniform shift so that
\begin{equation}
\label{eq:Omg1N2}
\omega_{2,m} = \omega_{1,m} + \frac{1}{T}, \quad \forall m \in [|\Omega_1|]. 
\end{equation}
Then it follows from the construction of $\bm{\Phi}$ that 
\begin{equation}
\label{eq:rotinv_Phi}
\bm \Pi_2 \bm \Phi = \bm \Pi_1 \bm \Phi \bm D
\end{equation}
where $\bm D \in \mathbb{C}^{S \times S}$ is a diagonal matrix satisfying $(\bm D)_{k,k} = e^{\frac{-\mathsf{j} 2 \pi \tau_k}{T}}$. 
Furthermore, if $\bm{G}$ satisfies the \textit{Pairwise-Identical Condition} (PIC) defined by \begin{equation}
\label{eq:pcc1}
\bm \Pi_1 \bm G \bm \Pi_1^\top = \bm \Pi_2 \bm G \bm \Pi_2^\top,
\end{equation}
then the rotation-invariance property in \eqref{eq:rotinv_GPhi} follows from \eqref{eq:rotinv_Phi} and \eqref{eq:pcc1}. 
The PIC in \eqref{eq:pcc1} is equivalently rewritten as 
\begin{equation}
\label{eq:pcc}
G(\omega_{1,m}) = G(\omega_{2,m}), \quad m \in [|\Omega_1|]. 
\end{equation}

In practice, $\widehat{\bm{U}}$ can provide only an estimate of the signal subspace. 
The exact PIC condition is too stringent for pulse shapes in practical applications. 
Therefore, it is of great interest to extend the analysis beyond the noiseless scenario with the exact PIC considered in the previous work \cite{bresler1989resolution}. 
The following proposition illustrates how the error, resulting from both the violation of the exact PIC condition and the presence of noise, propagates to the estimation of the pulse locations.
The accuracy of our estimated locations  
$\widehat{\mathcal{T}} := \{\widehat{\tau}_k\}_{k=1}^S$ are compared to the ground-truth locations $\mathcal{T} := \{\tau_k\}_{k=1}^S$ via the \textit{matching distance} metric \cite{li2020super} defined by
\begin{align*}
\mathrm{md}(\mathcal{T}, \widehat{\mathcal{T}}) 
:= 
\min_{\pi \in \mathrm{perm}(S)} \max_{k \in [S]} |\tau_k - {\widehat{\tau}_{\pi(k)}}|
\end{align*}
where $\mathrm{perm}(S)$ denotes the set of all possible permutations over $[S]$.

\begin{proposition}
\label{prop:gen_case_deterministic}
Let $\Omega$ satisfy $|\Omega| \geq S+1$ and $\Omega_1, \Omega_2$ be constructed as in \eqref{eq:Omg1N2}. Let $\bm U_1 = \bm \Pi_1 \bm U$ where $\bm U \in \mathbb{C}^{|\Omega| \times S}$ spans the column space of $\bm{G}\bm{\Phi}$ and satisfies $\bm U^\mathsf{H} \bm U = \bm I_S$. 
Suppose that $\bm X$ has full row rank and $\widehat{\bm U}$ satisfies 
\begin{align}
\label{eq:prop_cond}
\min_{\bm R \in O_S} \norm{\widehat{\bm U} - \bm{U} \bm R} < \frac{\sigma_S(\bm{U}_1)}{2}.
\end{align}
where $O_S$ denotes the set of $S $-by-$S$ unitary matrices. 
Then the estimated locations by ESPRIT satisfies
\begin{align}
\label{eq:det_err_bnd}
& \mathrm{md}(\mathcal{T},\widehat{\mathcal{T}}) 
\leq  \frac{\kappa({\bm{\Phi}}) G_{\max}}{2 \Gamma G_{\min}}
\nonumber \\
& \quad \cdot \left(\frac{3\, \mathrm{dist}(\bm{\widehat{U}},\bm U)}{\sigma_S^2(\bm{U}_1)}   +  \frac{\max_m |G(\omega_{2,m}) - G(\omega_{1,m})|}{\sigma_S(\bm{U}_1) G_{\min}} \right)
\end{align}
where
\begin{equation}
\label{eq:def_Gmin_Gmax}
G_{\min} := \min_{\omega \in \Omega} |G(\omega)|
\quad \text{and} \quad 
G_{\max} := \max_{\omega \in \Omega} |G(\omega)|.
\end{equation}
\end{proposition}

Proposition~\ref{prop:gen_case_deterministic} extends the non-asymptotic perturbation analysis of ESPRIT \cite{li2020super,li2022stability} to the unknown pulse shape case. 
Proposition~\ref{prop:gen_case_deterministic} explains how the violation of the condition in \eqref{eq:pcc} propagates to the error in estimating the pulse locations. 
In particular, the maximum deviation $\max_m |G(\omega_{2,m}) - G(\omega_{1,m})|$ quantifies the degree of the model error induced by the violation of the exact PIC in \eqref{eq:pcc}.

In the following sections, we specify Proposition~\ref{prop:gen_case_deterministic} to selected scenarios of the sampling pattern and the noise distribution so that the sufficient condition in \eqref{eq:prop_cond} and the error bound in \eqref{eq:det_err_bnd} are given as a simple function of the number of snapshots and the number of Fourier measurements per snapshot. 
For example, the condition number $\kappa({\bm{\Phi}})$ of the Vandermonde matrix $\bm{\Phi}$ is upper-bounded depending on the sampling pattern and minimum separation of the sources defined by
\begin{equation}
\label{eq:def_min_sep}
\Delta := \min_{k \neq j}|\tau_k - \tau_j|_{\mathbb{T}}.
\end{equation}
Moreover, in the presence of additive Gaussian noise to $\bm{Y}$, principal component analysis has been used to estimate the signal subspace. 
The classical perturbation theory of principal component analysis provides an upper bound on $\mathrm{dist}(\bm{\widehat{U}},\bm U)$ depending on the spectrum of the noise-free data matrix $\bm{Y}$. 

\section{ESPRIT with maximum overlapping sub-arrays} 
\label{sec:max_overlap}

The first scenario focuses on the case where the cost of snapshots is more expensive than the cost of Fourier measurements. 
For example, the signal model in \eqref{eq:signal_model_td} can describe snapshots of signals observed by electrodes in the surface electromyography (sEMG). 
The pulse in this context corresponds to the motor-unit action potential. 
Determining its shape is of great interest in various applications of sEMG. 
With a dense array of electrodes, sEMG provides a spatio-temporal view of neural signals. 
However, there exists a physical limitation on the density of an electrode array with the current fabrication techniques. 
It prevents increasing the number of snapshots $L$ above a certain threshold. 
The sEMG modality obtains measurements as time-domain samples typically at the rate of $2$ kHz \cite{chandra2020performance}. 
The signal in the form of \eqref{eq:signal_model_td} is not strictly band-limited, particularly with a narrow pulse. 
However, the anti-alias low-pass filter only affects the pulse shape due to the associativity of the convolution operator. 
Alternatively, one can use the acquisition with specially designed analog filters \cite{vetterli2002sampling,tur2011innovation}. 
As described earlier, this paper will consider the reconstruction from Fourier measurements. 
In the sEMG context, the number of Fourier measurements $M$ is determined by the sampling rate in the time domain and is typically a large number.

Furthermore, this scenario considers the additive white Gaussian noise (AWGN) to the data matrix $\bm{Y}$ in \eqref{eq:decomp_noisefree_data}. 
In other words, the observed noisy data is given by $\widehat{\bm{Y}} = \bm{Y} + \bm{Z} \in \mathbb{C}^{|\Omega| \times L}$ where the entries of $\bm{Z}$ are i.i.d. $\mathrm{Normal}(0,\sigma^2)$. 
Then the column space of 
$\widehat{\bm{Y}}$ is estimated using principal component analysis. 
Let $\widehat{\bm{U}} \in \mathbb{C}^{|\Omega| \times S}$ be a matrix whose columns corresponds the $S$-dominant eigenvectors of the empirical covariance matrix $\bm{R}_{\widehat{\bm{Y}}} = \frac{1}{L} \widehat{\bm{Y}} \widehat{\bm{Y}}^\mathsf{H}$. 
Note that this subspace estimation step requires $L \geq S$. The estimated ortho-basis by $\widehat{\bm{U}}$ is fed as an input to the ESPRIT algorithm.

Our goal here is to identify the number of snapshots $L$ providing a stable estimate of the pulse locations 
in the above scenario. 
Therefore, we explore the situation where $M$ is large, but $L$ carries an associated cost. 
We aim to answer this question in the context of the max-overlapping sub-arrays, a sampling pattern commonly used for the construction of sub-arrays in ESPRIT in the literature \cite{swindlehurst1999methods,astely1999spatial,asztely1997auto}. 

The index sets for the two sub-arrays $\Omega_1$ and $\Omega_2$ in the maximum overlapping case are constructed as
\begin{align}
\Omega_1 & = \left\{  \frac{l}{T} : l = 0,1,\dots,M -1 \right\} 
\quad \text{and} \quad \nonumber \\ 
\Omega_2 & = \left\{  \frac{l}{T} : l = 1,2, \dots,M \right\}.
\label{eq:Omega1_fs}
\end{align}

\begin{figure} [!ht]
		\begin{subfigure}
      {.45\textwidth}
			\centering
\includegraphics[width=.9\linewidth]{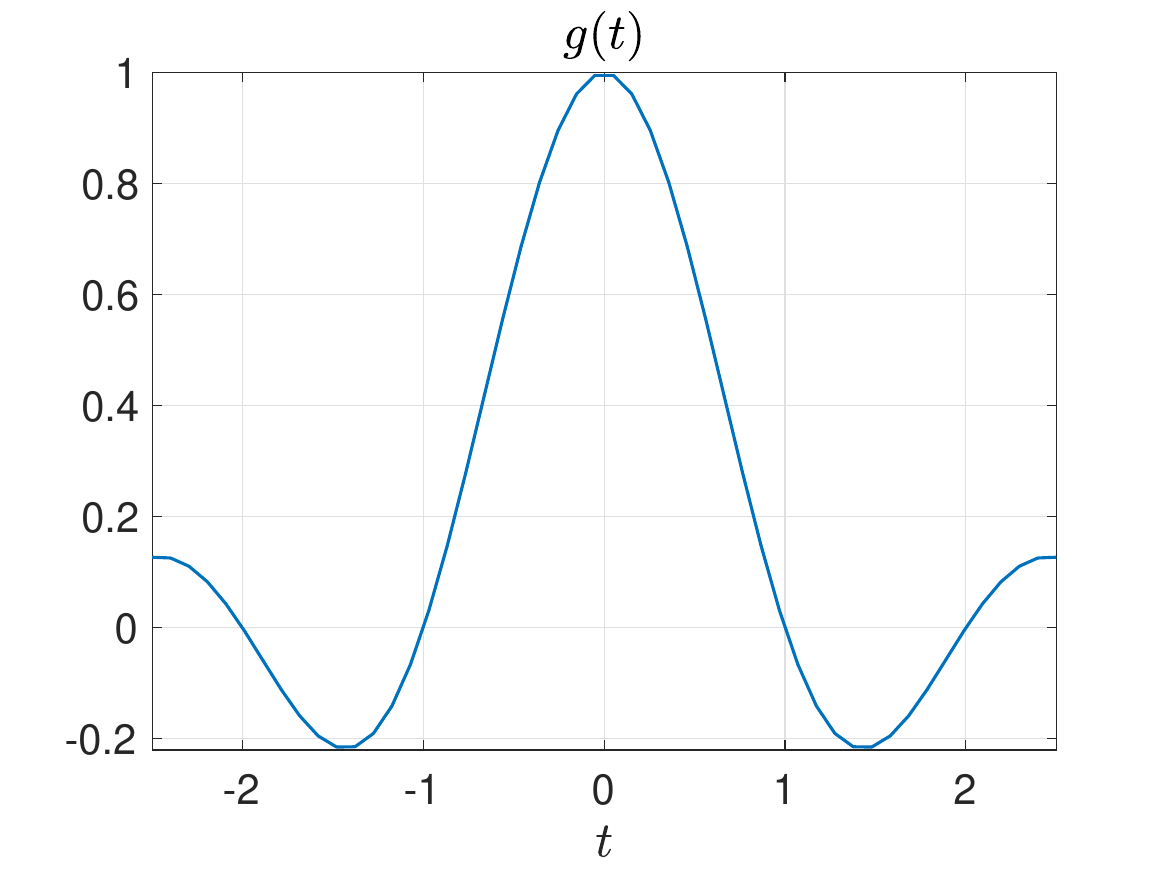}
 \label{fig:sinc pulse}
		\end{subfigure}
		\begin{subfigure}{.45\textwidth}
			\centering
\includegraphics[width=.9\linewidth]{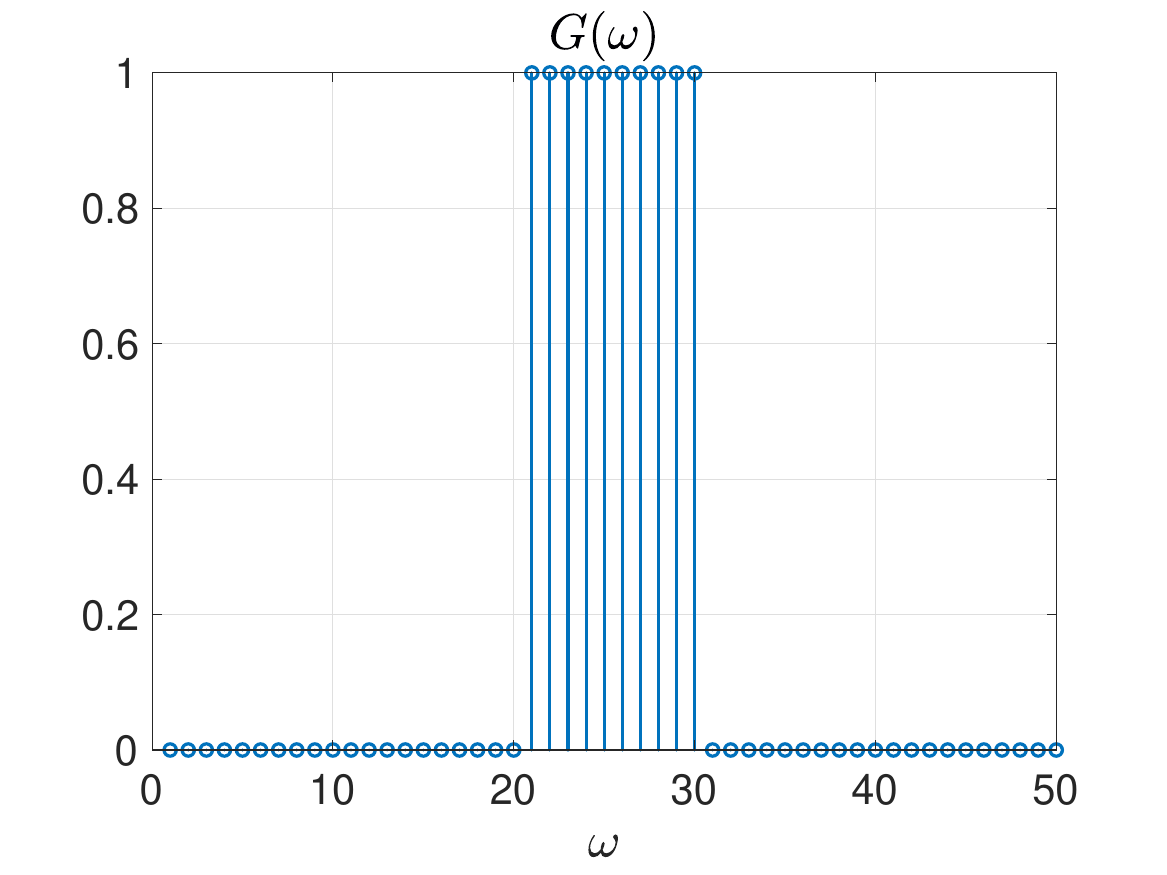}
   \label{fig:rect window}
		\end{subfigure}	
  \caption{The sinc pulse shape satisfies pairwise-identical condition exactly in its Fourier domain.}
  \label{fig:sinc_pulse}
\end{figure}

Then the PIC condition in \eqref{eq:pcc} reduces to 
\begin{equation}
\label{eq:pcc_max_overlap}
G\left(\frac{l}{T}\right) = G(0), \quad \forall l = 1,2, \dots,M.
\end{equation}
For example, if the pulse shape corresponds to a sinc function shown in Fig~\ref{fig:sinc_pulse}, then it satisfies \eqref{eq:pcc_max_overlap}. 

However, the pulse shapes in real-world applications does not exactly satisfy \eqref{eq:pcc_max_overlap}. 
Therefore, we want to carry out the perturbation analysis of ESPRIT to see how the violation of \eqref{eq:pcc_max_overlap} affects the estimation error. 
The following theorem presents the specification of the Proposition~\ref{prop:gen_case_deterministic} to the current scenario for ESPRIT with the maximum-overlapping sub-arrays. 

\begin{theorem}
\label{thm:fully_sampled_case} 
Let the hypothesis of the Proposition~\ref{prop:gen_case_deterministic} hold, $\Omega_1, \Omega_2$ be constructed as in \eqref{eq:Omega1_fs}, and $\widehat{\bm{U}}$ be an estimate of $\bm{U}$ from $\widehat{\bm{Y}} = \bm{Y} + \bm{Z}$ via principal component analysis, where $\bm{Y}$ is the noise-free data in \eqref{eq:decomp_noisefree_data} and 
the entries of $\bm{Z}$ are i.i.d. $\mathrm{Normal}(0,\sigma^2)$. 
Then there exist absolute constants $C_1,c_2>0$ such that 
\begin{equation}
\label{eq:thm:errbnd_fs}
\begin{aligned} 
& \mathrm{md}(\mathcal{T}, \widehat{\mathcal{T}}) 
\lesssim \frac{T \rho \sigma}{G_{\min} \lambda_S^{1/2}(\bm{R}_{\bm{X}}) \sqrt{L} \left(1-\frac{2\rho^2 S}{M}\right)}  \\
& \qquad \cdot \left(  \rho \kappa^{1/2}(\bm{R}_{\bm{X}}) \vee \frac{\sigma}{G_{\min} \lambda_S^{1/2}(\bm{R}_{\bm{X}}) \sqrt{M} } \right) \\
& \quad + \frac{\rho}{{\left(1-\frac{2\rho^2 S}{M}\right)}^{1/2}} \cdot \frac{\sup_{\omega \in \Omega_1} |G(\omega+\frac{1}{T}) - G(\omega)|}{\frac{1}{T} \cdot G_{\min}}
\end{aligned}
\end{equation}
holds with probability $1 - 3e^{-c_2M}$, provided that
\begin{equation}
\label{eq:thmcond_M_fs}
M \geq S \vee \left( \frac{3}{\Delta}+2 \right)
\end{equation}
and
\begin{equation}
\label{eq:thmcond_L_fs}
L \geq S \vee \frac{C_1 \sigma^2}{G^2_{\min} \lambda_S(\bm{R}_{\bm{X}})} \cdot \left( \frac{ \sigma^2}{M G^2_{\min} \lambda_S(\bm{R}_{\bm{X}})} \vee \frac{\rho^2 \kappa(\bm{R}_{\bm{X}})}{1-\frac{2\rho^2 S}{M}} \right).
\end{equation}
\end{theorem}

Theorem~\ref{thm:fully_sampled_case} considers a scenario where the number of Fourier measurements $M$ is sufficiently large relative to the minimum separation. 
The first term in the error bound in \eqref{eq:thm:errbnd_fs} illustrates the rate at which the estimation error due to noise decreases as $L$ increases. 
It also depends on the dynamic range of the pulse shape in the Fourier domain through the parameter $\rho$. 
For a fixed pulse shape and $M$, the first error term decays as $\frac{\sigma}{\sqrt{L}}$. 
However, the second error term in \eqref{eq:thm:errbnd_fs} prevents ESPRIT to be a consistent estimator even in the noiseless scenario. 
The second term is due to the model error from the violation of the PIC. 
If $g(t)$ is supported within the interval $[-R/2, R/2)$, by the mean value theorem, the second term in the error bound in \eqref{eq:thm:errbnd_fs} simplifies through an upper bound as
\[
\frac{|G(\omega+\frac{1}{T}) - G(\omega)|}{\frac{1}{T}} 
\leq \sup_{\omega \in \mathbb{R}} \left| \frac{d G(\omega)}{d\omega} \right| 
\leq 2\pi R \sup_{\omega \in \mathbb{R}} |G(\omega)|
\]
where the second inequality is due to Bernstein's inequality (cf. \cite[Theorem~6.7.1]{lapidoth2017foundation}). 
This implies that the error due to the violation of the exact PIC in  \eqref{eq:pcc} is small for narrow pulse shapes. 

\section{ESPRIT with random-doublet sub-arrays}
\label{sec:random_doublets}

The second scenario focuses on cases where the cost of a Fourier measurement is more crucial than the cost of snapshots. 
The data model in \eqref{eq:decomp_noisefree_data} also arises in a direction of arrival (DoA) estimation problem \cite{asztely1997auto,astely1999spatial,swindlehurst1999methods}. 
In the context of the DoA estimation, the parameters $M$ and $L$ denote the number of receivers in a sensor array and the number of time samples, respectively. 
Furthermore, each entry of the diagonal matrix $\bm{G}$ in \eqref{eq:decomp_noisefree_data} represents the unknown gain and phase of the corresponding receiver. 
The blind array calibration problem refers to the process of estimating this unknown calibration matrix $\bm{G}$ from a noisy version of data matrix $\bm{Y}$, without relying on known reference signals (pre-determined signals used for calibration).

The ultimate goal of the DoA estimation is to compute the steering vectors which correspond to the columns of $\bm{\Phi}$ in \eqref{eq:decomp_noisefree_data}.
If the exact calibration matrix $\bm{G}$ is known, then finding $\bm{\Phi}$ reduces to an easier harmonic retrieval problem. 
In the presence of noise, the stability of the estimation of $\bm{\Phi}$ crucially depends on its condition number. 
In a special case where $\bm{\Phi}$ is a Vandermonde matrix corresponding to a uniform linear array, the separation of pulses (i.e. the separation of sources in the DoA estimation) determines the required number of Fourier measurements (the number of receivers in DoA) for a stable estimation of $\bm{\Phi}$ \cite{moitra2015super}. 
It has been shown that if $M > \frac{1}{\Delta} + 1$, then the condition number of $\bm{\Phi}$ is upper-bounded by
\[
\kappa(\bm \Phi)  \leq \sqrt{\frac{M+1/\Delta-1}{M-1/\Delta-1}}.
\]
Furthermore, if $M \leq \frac{1-\epsilon}{\Delta}$ for any $\epsilon \in (0,1)$, there is no estimator that can recover all $\Delta$-separated pulses.\footnote{A subsequent result refined the analysis on special cases where the pulses are supported in specific structures \cite{li2020super}. Since we propose to lower the burden on $M$ in this paper, we still consider the worst-case instance of $\Delta$-separated signals.} 
Therefore, it is necessary to implement the system with a large number of receivers to resolve closely located sources when a uniform linear array is considered. 
The condition on $M$ is also required in a more challenging case with unknown $\bm{G}$. 

We propose to resolve closely located pulses of unknown shape by ESPRIT with random sub-arrays. 
Recall that the ESPRIT algorithm in Section~\ref{sec:max_overlap} can operate with a flexible design of sub-arrays without being restricted to the uniform linear structure. 
We consider sub-arrays obtained as random subsets of a large uniform linear array such that the two sub-arrays are related through a uniform shift. 

A similar idea has been proposed for the compressed sensing off-the-grid where the pulse shape is an impulse \cite{tang2013compressed}. 
Our design is summarized as follows: 
Let $\widetilde{\Omega}$ correspond to a uniform grid of size $2\widetilde{M}$ separated by $\frac{1}{2T}$ given by
\begin{equation}
\label{Omega_tilda_construction}
\widetilde{\Omega} := \left\{  \frac{l}{2T} : l = 0,1,\dots,2M -2 \right\}. 
\end{equation}
The first sub-array $\Omega_1$ is constructed as a random subset of $\widetilde{\Omega}$ given by
\begin{equation}
\label{eq:Omega1_random_doublets}
\Omega_1 = \left\{ \frac{2(i-1)}{2T} : i \in [\widetilde{M}], ~ \beta_i = 1 \right\}
\end{equation}
where $\beta_1 , \beta_2 \cdots , \beta_{\widetilde{M}/2}$ are independent copies of a Bernoulli random variable $\beta$ satisfying $\mathbb{P}(\beta = 1) = M/\widetilde{M}$ and $\mathbb{P}(\beta = 0) = 1 - M/\widetilde{M}$. 
The second sub-array $\Omega_2$ is then determined by $\Omega_1$ as in \eqref{eq:Omg1N2}. 

The following theorem presents the specification of the Proposition~\ref{prop:gen_case_deterministic} to the above scenario by utilizing the ESPRIT with the randomly sampled sub-arrays.

\begin{theorem} 
\label{thm:main}
Let the hypothesis of Theorem~\ref{thm:fully_sampled_case} holds except $\Omega_1, \Omega_2$ be constructed as in \eqref{eq:Omega1_random_doublets} from
$\widetilde{\Omega}$ given in \eqref{Omega_tilda_construction}. 
Then there exist absolute constants $C_1,C_2,c_3 > 0$ such that if 
\begin{equation}
\label{eq:thmcond_M_tilde}
    \widetilde{M} > S \vee 3 \left( \frac{1}{\Delta} + 1 \right),
\end{equation}
\begin{equation}
\label{eq:thmcond_M}
M \geq C_1 S \ln M,
\end{equation}
and
\begin{equation}
\label{eq:thmcond_L}
L \geq S \vee \frac{C_2 \sigma^2 \widetilde{\rho}^2 }{\widetilde{G}^2_{\min} \lambda_S(\bm{R}_{\bm{X}})} \left( \widetilde{\rho}^2 \kappa(\bm{R}_{\bm{X}}) \vee \frac{ \sigma^2}{M\widetilde{G}^2_{\min} \lambda_S(\bm{R}_{\bm{X}})}  \right),
\end{equation}
then it holds with probability $1 - 5e^{-c_3M} - M^{-1}$ that
\begin{equation}
\label{eq:thm:errbnd}
\begin{aligned}
& \mathrm{md}(\mathcal{T},\widehat{\mathcal{T}}) 
\lesssim \frac{T \widetilde{\rho}^3 \sigma}{\widetilde{G}_{\min} \lambda_S^{1/2}(\bm{R}_{\bm{X}}) \sqrt{L}} \\ 
& \qquad \cdot \left(  \widetilde{\rho} \kappa^{1/2}(\bm{R}_{\bm{X}}) \vee \frac{\sigma}{\widetilde{G}_{\min} \lambda_S^{1/2}(\bm{R}_{\bm{X}}) \sqrt{M} } \right) \\
& \quad + \widetilde{\rho}^2 \cdot \frac{\sup_{\omega \in \widetilde{\Omega}} |G(\omega+\frac{1}{2T}) - G(\omega)|}{\frac{1}{2T} \cdot \widetilde{G}_{\min}}, 
\end{aligned}
\end{equation}
where
\begin{equation}
\label{eq:def_Gmin_Gmax_tilde}
\widetilde{G}_{\min} := \min_{\omega \in \widetilde{\Omega}} |G(\omega)|, ~
\widetilde{G}_{\max} := \max_{\omega \in \widetilde{\Omega}} |G(\omega)|, ~
\widetilde{\rho} = \frac{\widetilde{G}_{\max}}{\widetilde{G}_{\min}}. 
\end{equation}
\end{theorem}

\begin{remark}
The requirement on $\widetilde{M} > 3 \left(\frac{1}{\Delta} + 1 \right)$ in \eqref{eq:thmcond_M_tilde} is introduced to simplify the proof. It can be reduced arbitrarily close to $\widetilde{M} > \frac{1}{\Delta} + 1$. 
\end{remark}

Similar to the analysis of Theorem~\ref{thm:fully_sampled_case}, the first term in the error bound in \eqref{eq:thm:errbnd} is due to noise and decays as $\frac{\sigma}{\sqrt{L}}$ when $\widetilde{M}$ is sufficiently large.
However, increasing $\widetilde{M}$ is accompanied with a penalty as $\widetilde{\rho}$ scales in accordance with the increasing dynamic range between $\widetilde{G}_{\max}$  and $\widetilde{G}_{\min}$, and the error increases proportionally to $\widetilde{\rho}$. 
For closely located pulses with small $\Delta$, the requirement on $\widetilde{M}$ in \eqref{eq:thmcond_M_tilde} can be stringent. 
However, the cost of the system is determined not by $\widetilde{M}$ but by $M$. 
The requirement on $M$ in \eqref{eq:thmcond_M} scales only proportional to $S$ up to a logarithmic factor.  
The second term in \eqref{eq:thm:errbnd} is due to the violation of the PIC. 
For example, the degree of violation decreases as the width of the pulse shape in the time domain becomes narrower.

\section{Numerical Results}
\label{sec:num_res}
We present numerical results that illustrate the empirical performance of ESPRIT in the two scenarios investigated in Sections~\ref{sec:max_overlap} and \ref{sec:random_doublets}. 
We show that these numerical results corroborate the theoretical results. 
Throughout the experiments, the amplitudes and noise entries are generated so that their entries are independent copies of a zero-mean Gaussian random variable. 
The Monte Carlo simulations compute statistics averaged on random noise and amplitudes while the pulse shape and locations are fixed. 
For the ESPRIT with random sub-arrays, sampling patterns will vary over instances in the Monte Carlo simulations. 

\begin{figure}[!htb]
    \centering
    \begin{subfigure}{0.32\textwidth}
   \centering
    \includegraphics[scale=0.33]{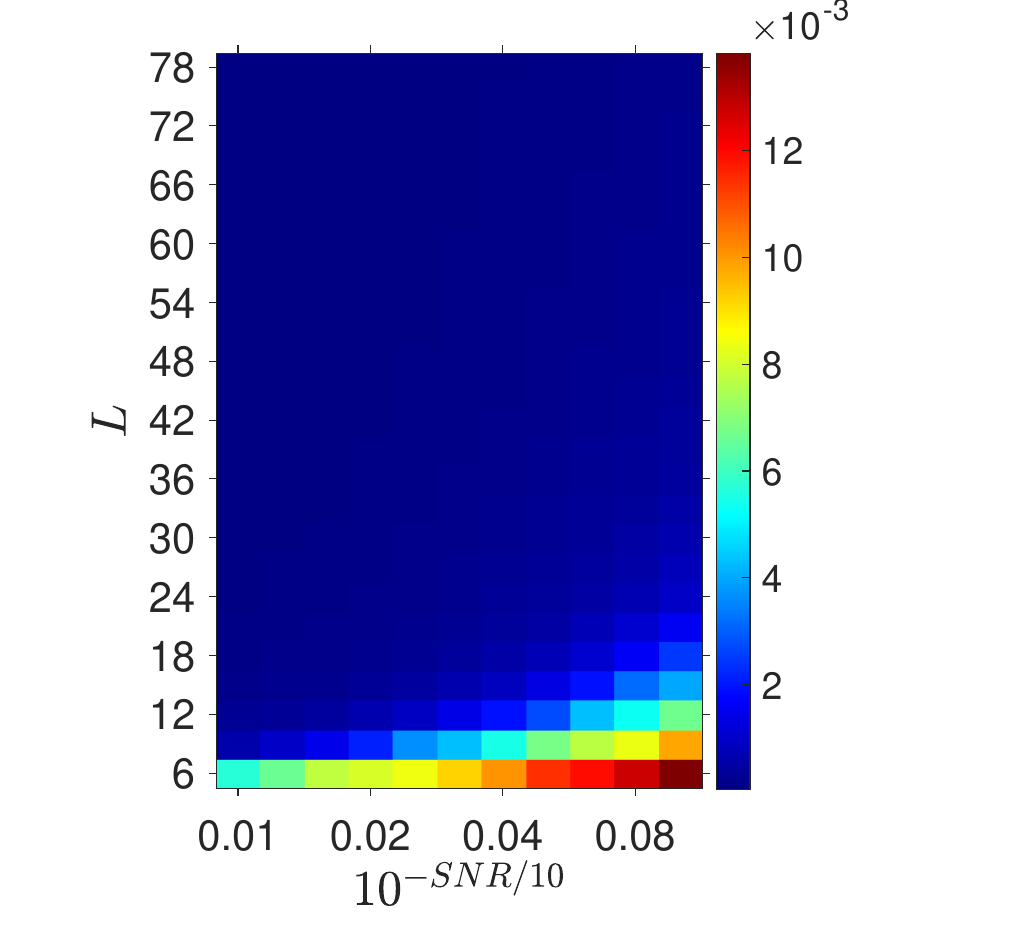}
	\caption{trivial pulse}
    \label{fig:phase_transition_triv}
   \end{subfigure}
    \begin{subfigure}{0.32\textwidth}
    \centering
    \includegraphics[scale=0.33]{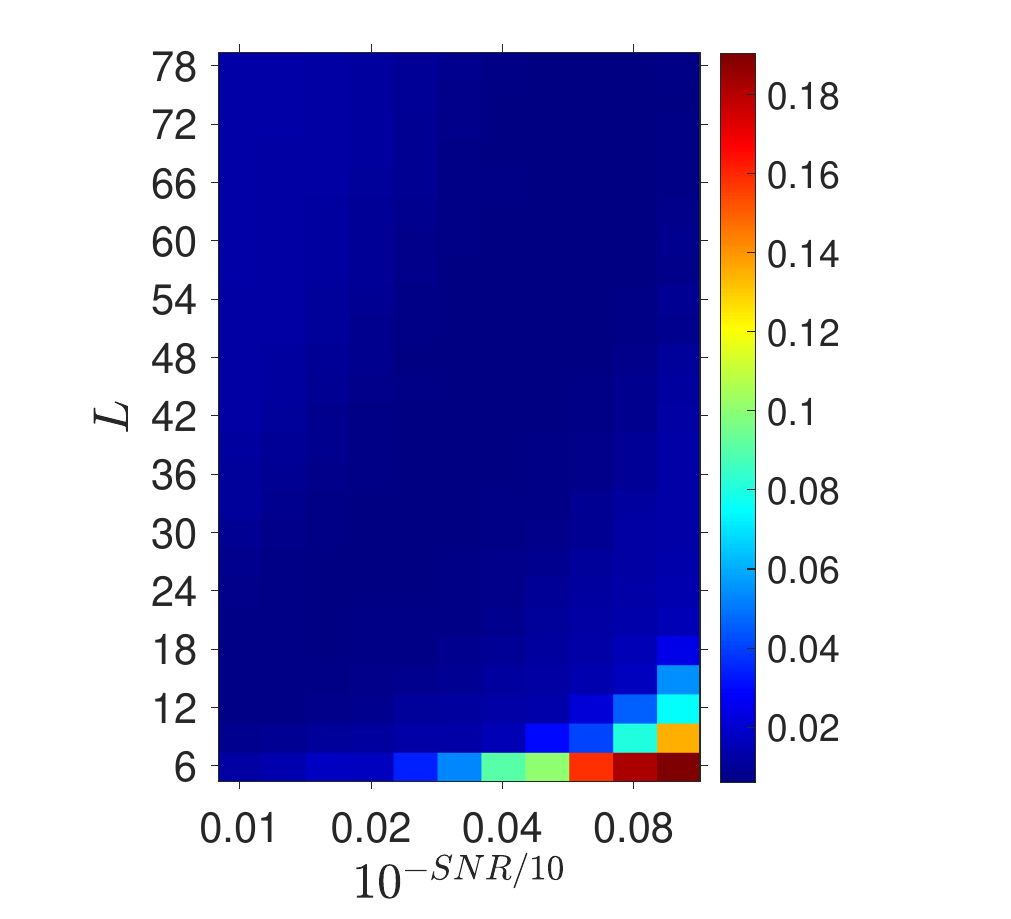}
	\caption{narrow pulse}
    \label{fig:phase_transition_narrow}
    \end{subfigure}
    \begin{subfigure}{0.32\textwidth}
    \centering
    \includegraphics[scale=0.33]{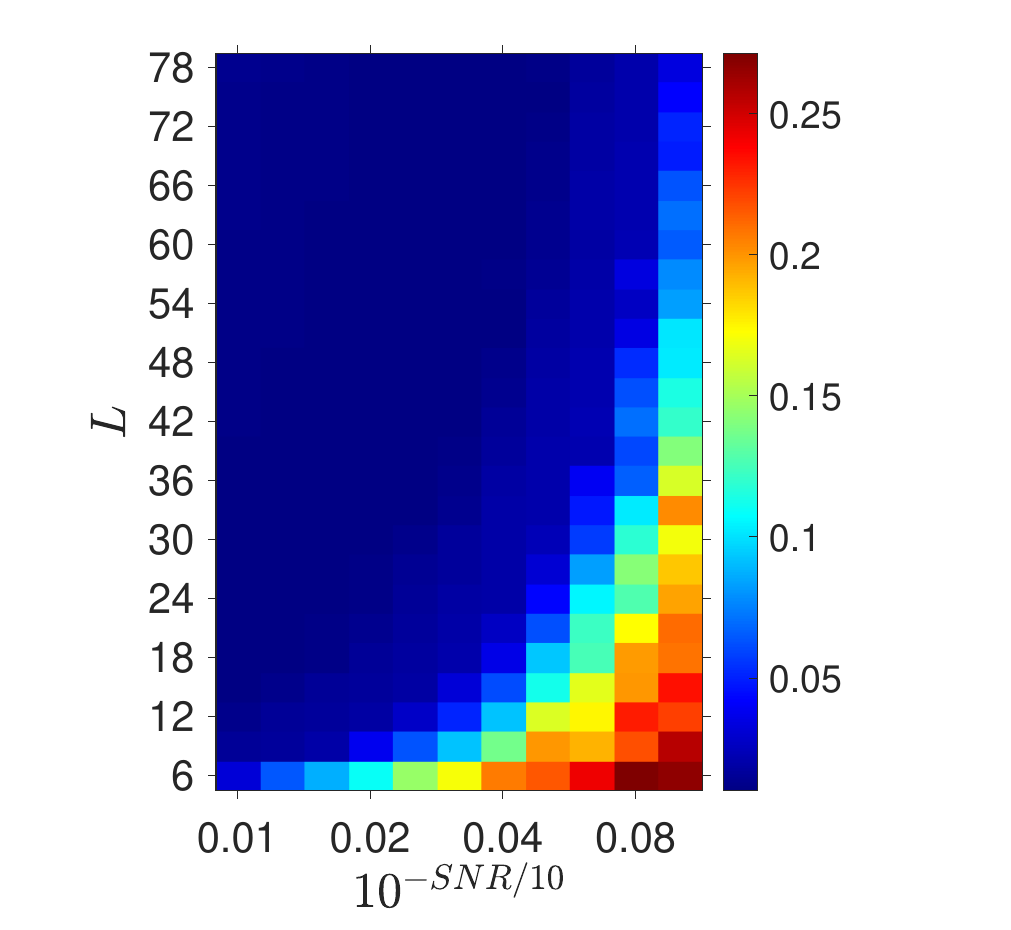}
    \caption{wide pulse}
    \label{fig:phase_transition_wide}
    \end{subfigure}	
    \caption{Phase transition for $L$ versus the relative noise $10^{\frac{-SNR}{10}}$ for various pulse shapes ($M = 240$, $\Delta = 0.0125$, $S=5$, SNR $\in \{10,20\}$)}
    \label{fig:phase_transition}
\end{figure}

\begin{figure}[!htb]
\centering
\includegraphics[scale=0.33]{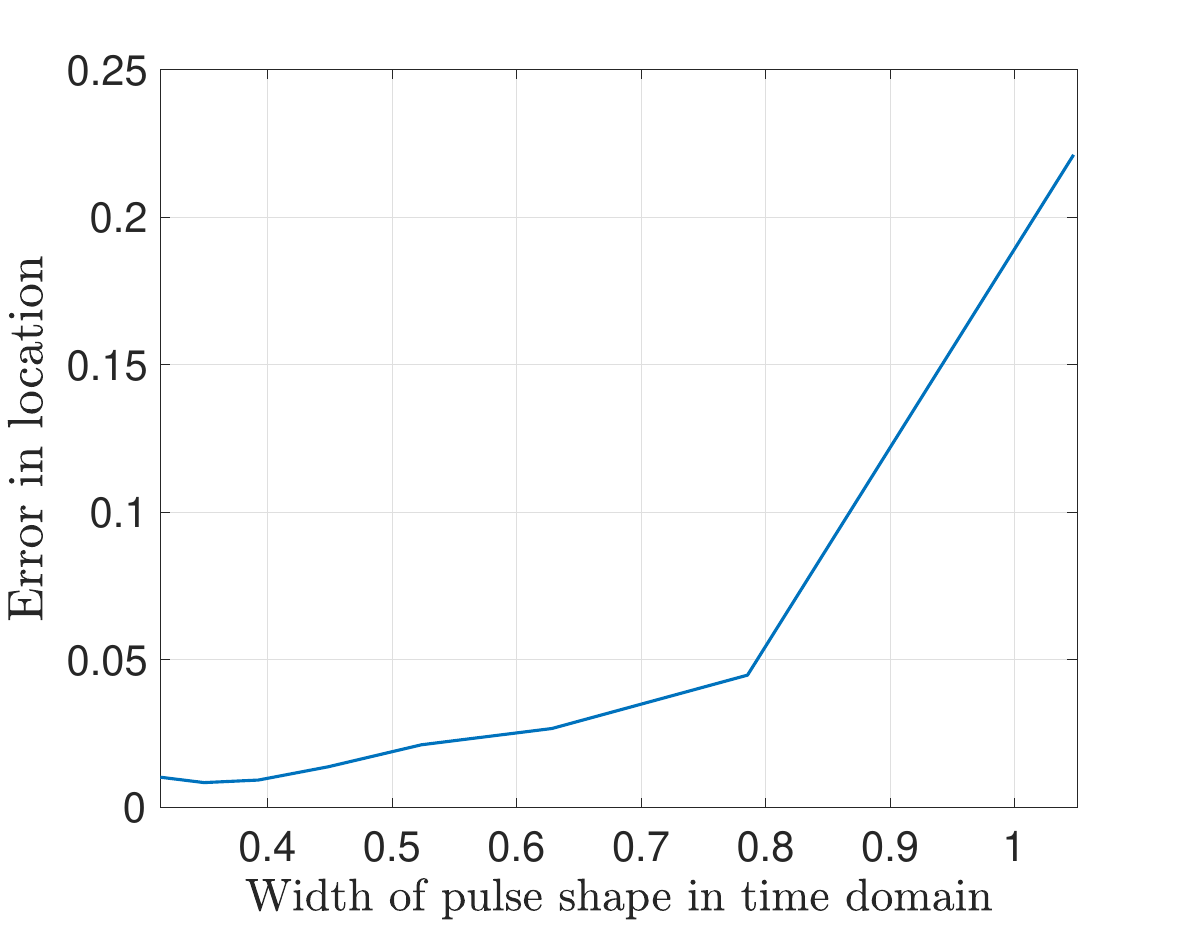}
\caption{Estimation error vs. pulse width (SNR $= 20$ db, $M = 160$, $\Delta = 0.025$, $S=5$, $L=20$)}{}
\label{fig:width_vs_error}
\end{figure}

We first observe the behavior of the estimator in the scenario in Sections~\ref{sec:max_overlap}. 
Fig.~\ref{fig:phase_transition} illustrate the empirical phase transition of the estimation error for $L$ versus the relative noise level $10^{\frac{-SNR}{10}}$, which is proportional to the noise variance $\sigma^2$. 
The experiment considers the trivial pulse shape $g(t) = \delta(t)$ and non-trivial pulse shapes in the form of $g(t) = \mathrm{rect}(20at/\pi)\cos^2(20at)$ for $a \in \{0.75,0.9\}$. 
The parameter $a > 0$ controls the pulse width (larger $a$ corresponds to a narrower pulse). 
In the trivial pulse case, the transition boundary occurs on a line given by the linear relation between $L$ and the relative noise level. 
A similar trend is observed for the narrow pulse case $(a=0.9)$. 
However, the wide pulse case $(a=0.75)$ shows a quadratic dependence of $L$ on the relative noise level at the transition boundary. 
The error in the non-trivial pulse case has a bias term that increases with the pulse width. This is due to the model error induced by the violation of the PIC. 
We also conduct an experiment that studies the effect of the pulse width on the estimation error when the width grows continuously. Fig.~\ref{fig:width_vs_error} shows the increase in the estimation error as the pulse width increases. This is because a wider pulse in the time domain leads to a greater violation of the PIC in the Fourier domain.

\begin{figure}[!htb]
\centering
\includegraphics[scale=0.33]{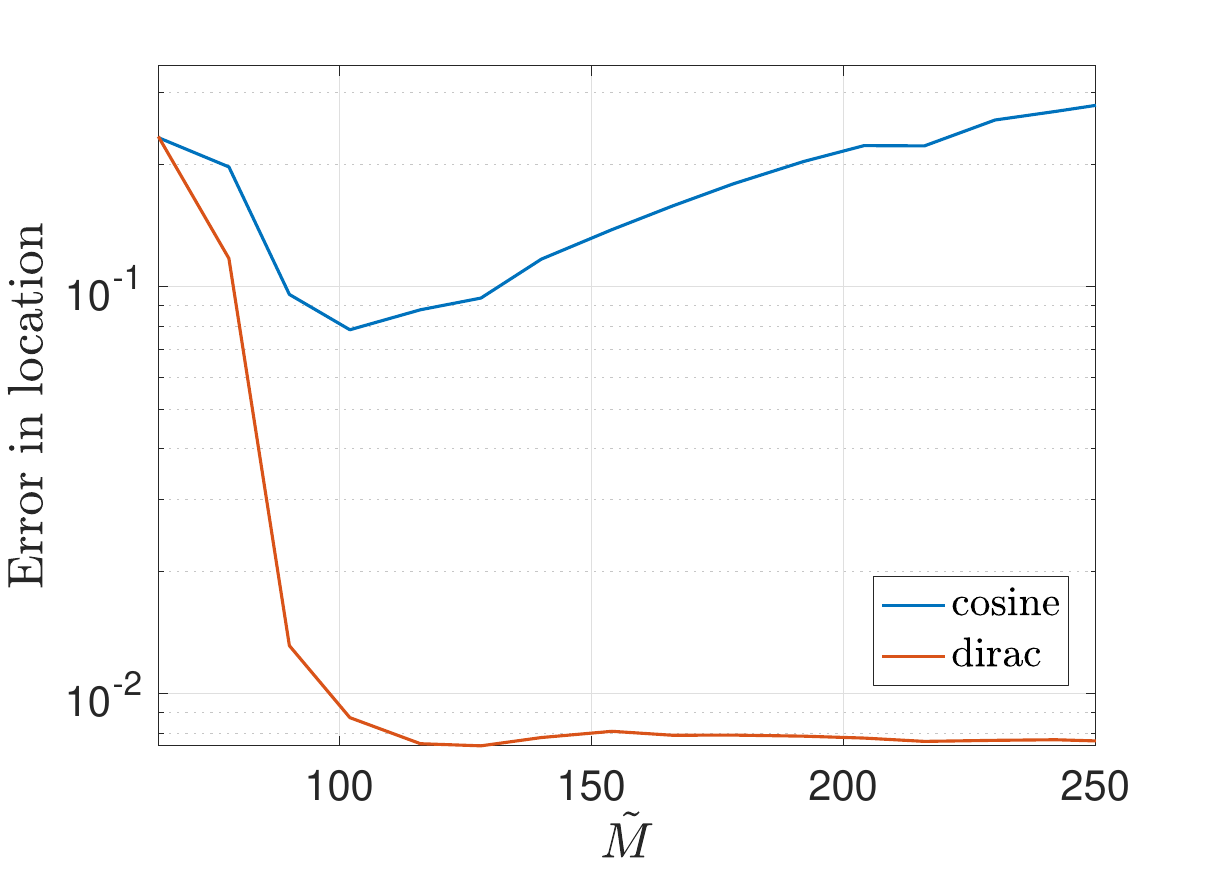}
\caption{Estimation error for $g(t) = \delta(t)$ and  $g(t) = \mathrm{rect}(20t/\pi)\cos^2(20t)$ (SNR $= 28$ db, $M = 40$, $\Delta = 0.008$, $S=7$, $L=50$). }{}
\label{fig:esti_err_for_two_pulse_shapes}
\end{figure}

\begin{figure}[!htb]
\centering
\includegraphics[scale=0.33]{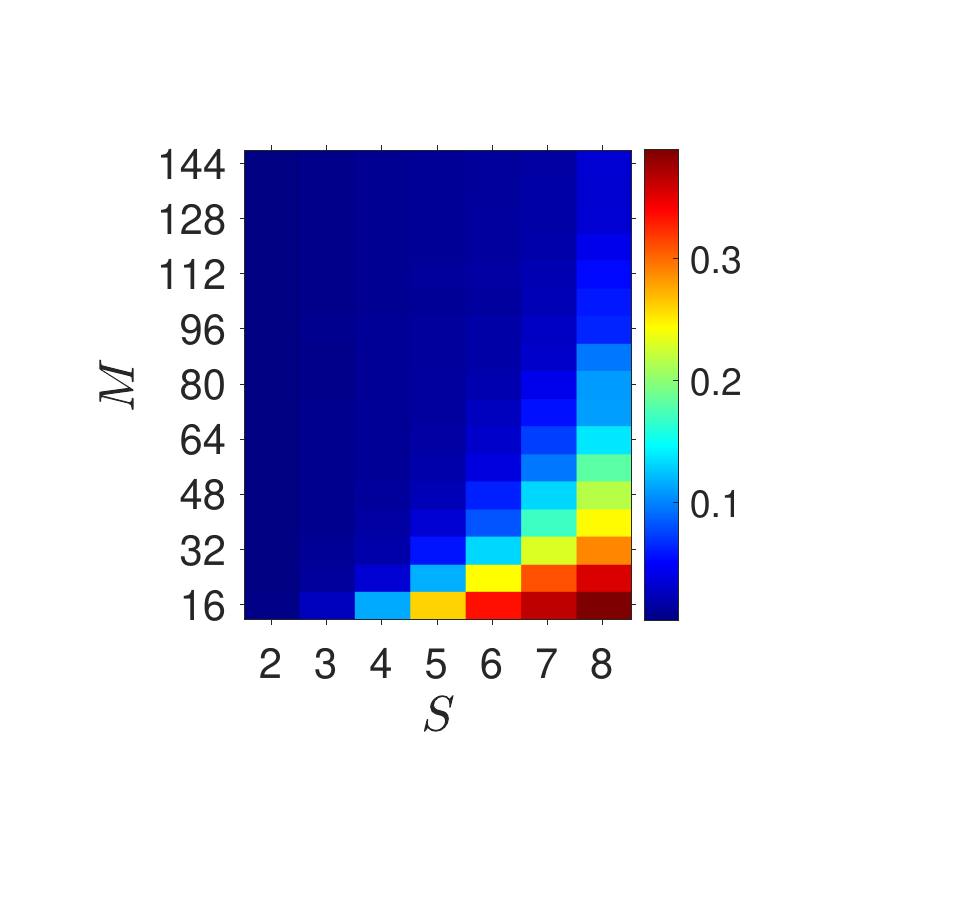}
\vspace{-7mm}
\caption{Empirical phase transition of estimation error for $g(t) = \mathrm{rect}(20t/\pi)\cos^2(20t)$ (SNR $= 26$ db, $\widetilde{M} = 260$, $\Delta = 0.008$, $L=50$).}
\label{fig:phase_trans_non_trivial}
\end{figure}

Next, we present the numerical results on the effectiveness of ESPRIT with the random sub-arrays in Section~\ref{sec:random_doublets}. 
First, we study how the choice of $\widetilde{M}$ affects the estimation error for the trivial (Dirac) and the non-trivial (truncated and squared cosine) pulse shapes. 
Fig.~\ref{fig:esti_err_for_two_pulse_shapes} illustrates the median of the estimation error from $1,000$ Monte Carlo trials for varying $\widetilde{M}$. 
In the trivial pulse shape case, the error saturates to a small value after $\widetilde{M}$ increases above a threshold ($\widetilde{M} \geq 120$). 
In this case, indeed the PIC in \eqref{eq:pcc} is satisfied and the parameters $\widetilde{G}_{\min}$, $\widetilde{G}_{\max}$, and $\tilde{\rho}$ in Theorem~\ref{thm:main} become $1$, hence the error bound in \eqref{eq:thm:errbnd} is invariant for all $\widetilde{M}$ satisfying \eqref{eq:thmcond_M_tilde}. 
In the non-trivial pulse shape case, the model error term in \eqref{eq:thm:errbnd} predominates due to the violation of the PIC and relatively weak noise level. 
Furthermore, the error remains high in the regime where $\widetilde{M}$ is not sufficiently large relative to the minimum separation as in \eqref{eq:thmcond_M_tilde}. 
When \eqref{eq:thmcond_M_tilde} is satisfied, further increasing $\widetilde{M}$ results in a higher dynamic range of the Fourier magnitudes of the pulse shape $|G(\omega)|$. 
Consequently, the error bound increases with larger values of the parameters $1/\widetilde{G}_{\min}$, $\widetilde{G}_{\max}$, and $\tilde{\rho}$. 
Next, Fig.~\ref{fig:phase_trans_non_trivial} the empirical phase transition of the estimation error for the number of pulses $S$ versus the number of Fourier measurements $M$. 
The median error over $500$ Monte Carlo trials is plotted. 
We observe that the relation between $M$ and $S$ on the transition boundary behaves similarly to \eqref{eq:thmcond_M}. 
Therefore, we conclude that these empirical observations bolster the theoretical result given in Theorem~\ref{thm:fully_sampled_case} and Theorem~\ref{thm:main}.

\section{Conclusion and Discussions}
\label{sec:conclusion}

This paper established a unifying perturbation analysis of resolving pulses of an unknown shape from multiple snapshots via ESPRIT that applies to any design of sub-arrays related by a uniform shift. 
The performance guarantee on the estimation error has been derived under a mild assumption that the pulse shape in the Fourier domain decays slowly. 
We quantified how the degree of the violation of the PIC and the noise propagates to the estimation error. 
This result overcomes the limitation of the existing theory relying on the stringent assumption on the signal model. 
The utility of the perturbation analysis has been illustrated over two practical scenarios in the context of surface electromyography and compressive blind array calibration. 
For the latter scenario, we propose a novel random sub-array design which effectively reduces the cost associated with Fourier measurements in array calibration applications. Our theoretical findings are supported by numerical results and offer a practical advancement in resolving overlapping pulses from noisy measurements.

The current analysis focuses on the resolution from Fourier measurements. 
When the pulse shape is the Dirac impulse distribution or known, the localization of pulses from filtered time samples have been studied in the literature \cite{vetterli2002sampling,mulleti2017paley,gedalyahu2011multichannel}. 
It would be of interest to extend the theoretical results in this paper to the resolution from filtered time samples.

\section{Acknowledgements}
\noindent The authors would like to extend their heartfelt gratitude to Yoram Bresler for his insightful discussions in this work. 
\appendix
\subsection{Proof of Proposition \ref{prop:gen_case_deterministic}}

Recall that the ESPRIT algorithm estimates the pulse locations as the eigenvalues of $\widehat{\bm \Psi} = \widehat{\bm U}_1^\dagger \widehat{\bm U}_2$ where $\widehat{\bm U}_1 = \bm \Pi_1 \widehat{\bm U}$ and $\widehat{\bm U}_2 = \bm \Pi_2 \widehat{\bm U}$. 
Since $\bm U$ spans the column space of $\bm Y = \bm G \bm \Phi \bm X$ and $\bm X$ has full row rank, there exists an invertible matrix $\bm P \in \mathbb{C}^{S \times S}$ such that $\bm U = \bm G \bm \Phi \bm P$.

Suppose that the oracle information of $\bm U_1 = \bm \Pi_1 \bm U = \bm \Pi_1 \bm G \bm \Pi_1^\top \bm \Pi_1 \bm \Phi \bm P$ and $\breve{\bm U}_2 = \bm \Pi_1 \bm G \bm \Pi_1^\top \bm \Pi_2 \bm \Phi \bm P$ is available. 
By the construction of $\bm \Pi_1$ and $\bm \Pi_2$, it follows that $\bm \Pi_1 \bm \Phi$ and $\bm \Pi_2 \bm \Phi$ satisfy \eqref{eq:rotinv_Phi}. 
Therefore, applying the ESPRIT algorithm to $\bm U_1$ and $\breve{\bm U}_2$ reconstructs the ground-truth $\tau_1,\dots,\tau_S$ from the eigenvalues of $\breve{\bm \Psi} := \bm{U}_1^\dagger \breve{\bm U}_2$. 

Let $\bm R \in \mathbb{C}^{S \times S}$ be an arbitrary rotation matrix satisfying $\bm R^\mathsf{H} \bm R = \bm I_S$. Note that $\bm U \bm R$ and $\bm U$ share the same column space. 
They also satisfy $\mathrm{dist}_2(\bm{\widehat{U}},\bm U \bm R) = \mathrm{dist}_\mathrm{F}(\bm{\widehat{U}},\bm U)$ and $\mathrm{dist}_\mathrm{F}(\bm{\widehat{U}},\bm U \bm R) = \mathrm{dist}_\mathrm{F}(\bm{\widehat{U}},\bm U)$. 
Furthermore, since $\bm R^{-1} \breve{\bm \Psi} \bm R = (\bm U_1 \bm R)^\dagger \breve{\bm U}_2 \bm R$ and $\breve{\bm \Psi}$ have the same eigenvalues, substituting $\bm U$ by $\bm U \bm R$ does not affect the estimation error. 
Therefore, without loss of generality, we may assume that $\norm{\widehat{\bm U} - \bm{U}} = \min_{\bm R \in O_S} \norm{\widehat{\bm U} - \bm{U} \bm R}$.

Next we derive an upper bound on the matching $\ell_\infty$ distance between the sets of eigenvalues of $\breve{\bm \Psi}$ and $\widehat{\bm \Psi}$. 
Then we show that it implies the corresponding matching distance between the pulse locations and their estimates. 

Since $\breve{\bm \Psi} = \bm U_1^\dagger \breve{\bm U}_2 = \bm{P}^{-1} \bm {D} \bm{P}$, by the Bauer-Fike theorem \cite[Theorem~IIIa]{bauer1960norms}, we have
\begin{align*}
\mathrm{md}(\Lambda(\breve{\bm{\Psi}}), \Lambda(\widehat{\bm{\Psi}})) 
\leq 
\kappa(\bm P) \norm{\widehat{\bm{\Psi}}-\breve{\bm{\Psi}}}.
\end{align*}
We derive an upper bound on the distance between $\breve{\bm \Psi}$ and $\widehat{\bm \Psi}$ by the triangle inequality
\begin{equation}
\label{eq:trieq_Psi}
\norm{\breve{\bm \Psi} - \widehat{\bm \Psi}} 
\leq \norm{\bm \Psi - \widehat{\bm \Psi}} + \norm{\breve{\bm \Psi} - \bm \Psi} 
\end{equation}
where $\bm \Psi := \bm U_1^\dagger \bm U_2$ and $\bm U_2 = \bm \Pi_2 \bm U$. 

It has been shown \cite{li2020super} that the first summand in the right-hand side of \eqref{eq:trieq_Psi} is upper-bounded by
\begin{equation}
\label{eq:smd_Psihat}
\norm{\widehat{\bm{\Psi}}-\bm\Psi} 
\leq \frac{3\sqrt{2} \, \mathrm{dist}(\bm{\widehat{U}},\bm U)}{\sigma_S^2(\bm{U}_1)}. 
\end{equation}
For the sake of completeness, we present its proof below. 
\begin{IEEEproof}[Proof of \eqref{eq:smd_Psihat}]
By the triangle inequality, we have
\begin{align}
\norm{\widehat{\bm{\Psi}}-\bm\Psi} 
& = \norm{\widehat{\bm U}_1^\dagger \widehat{\bm U}_2 - \bm{U}_1^\dagger \bm{U}_2} \nonumber \\
& = \norm{\widehat{\bm U}_1^\dagger \widehat{\bm U}_2 - {\bm{U}_1^\dagger} \widehat{\bm U}_2 + \bm{U}_1^\dagger \widehat{\bm U}_2 - \bm{U}_1^\dagger \bm{U}_2} \nonumber \\
& = \norm{(\widehat{\bm U}_1^\dagger - \bm{U}_1^\dagger ) \widehat{\bm U}_2 + \bm{U}_1^\dagger (\widehat{\bm U}_2 - \bm{U}_2)} \nonumber \\ 
& \leq \norm{\widehat{\bm U}_1^\dagger - \bm{U}_1^\dagger} \cdot \norm{\widehat{\bm U}_2}  + \norm{\bm{U}_1^\dagger} \cdot \norm{\widehat{\bm U}_2 -  \bm{U}_2} \nonumber \\
& \leq \underbrace{\norm{\widehat{\bm U}_1^\dagger - \bm{U}_1^{\dagger}}}_{(\mathrm{a})} + \norm{\bm{U}_1 ^\dagger}_2 \cdot \norm{\widehat{\bm U}_2 -  \bm{U}_2}. \label{eq:ub_dist_Psi_Psihat}
\end{align}
The summand (a) in \eqref{eq:ub_dist_Psi_Psihat} is upper-bounded by \cite[Theorem~4.1]{wedin1973perturbation} as
\begin{align*}
\norm{{\widehat{\bm U}_1^\dagger - \bm{U}_1^\dagger}} 
& \leq \norm{{\bm{U}_1^\dagger}} \cdot \norm{\widehat{\bm U}_1^\dagger} \cdot  \norm{\widehat{\bm{U}} - \bm{U}}.
\end{align*}
Furthermore, by \cite[Lemma~3.1]{wedin1973perturbation}, we have
\[
\norm{\widehat{\bm U}_1^\dagger}
\leq 
\frac{\norm{\bm{U}_1^\dagger}}{1-\norm{\bm{U}_1^\dagger} \cdot \norm{\widehat{\bm U}_1 -  \bm{U}_1}}.
\]
Due to the assumption that $\norm{\widehat{\bm U} - \bm U} \leq \sigma_S(\bm U_1)/2$, we have
\[
\norm{\bm{U}_1^\dagger} \cdot \norm{\widehat{\bm U}_1 - \bm{U}_1} 
\leq \sigma_S(\bm U_1)^{-1} \, \norm{\widehat{\bm U} - \bm{U}} \leq \frac{1}{2}. 
\]
Plugging in these results to \eqref{eq:ub_dist_Psi_Psihat} yields
\begin{equation}
\label{eq:Psihat_Fnorm}
\norm{\widehat{\bm{\Psi}}-\bm\Psi} 
\leq \left( 2 \norm{\bm{U}_1^\dagger}^2 + \norm{\bm{U}_1^\dagger} \right) \norm{\widehat{\bm{U}} - \bm{U}} 
\leq \frac{3 \norm{\widehat{\bm{U}} - \bm{U}}}{\sigma_S^2(\bm U_1)}.
\end{equation}
Since $\norm{\widehat{\bm{U}} - \bm{U}} \leq \sqrt{2} \mathrm{dist}(\bm{\widehat{U}},\bm U)$ \cite[Lemma~2.6]{chen2021spectral}, the assertion follows from \eqref{eq:Psihat_Fnorm}. 
\end{IEEEproof}  

The second summand in the right-hand side of \eqref{eq:trieq_Psi} is upper-bounded by
\begin{equation}
\label{eq:smd_Psibreve}
\norm{\breve{\bm{\Psi}}-\bm{\Psi}} \leq \frac{\max_{m \in [|\Omega_1|]} |G(\omega_{2,m}) - G(\omega_{1,m})|} {\sigma_S(\bm{U}_0) G_{\min}}.
\end{equation}

\begin{IEEEproof}[Proof of \eqref{eq:smd_Psibreve}]
For brevity, we introduce shorthand notations given by
\[
\bm{G}_j = \bm{\Pi}_j \bm{G} \bm{\Pi}_j^\top, \quad \bm{\Phi}_j = \bm{\Pi}_j \bm{\Phi}, \quad j = 1,2, 
\]
so that $\bm U_2$ and $\breve{\bm U}_2$ are respectively written as 
\[
\bm U_2 = \bm{G}_2 \bm \Phi_2 \bm{P} 
\quad \text{and} \quad
\breve{\bm U}_2 = \bm{G}_1 \bm \Phi_2 \bm{P}. 
\]
Then it follows from the definitions of $\breve{\bm \Psi} = \bm U_1^\dagger \breve{\bm U}_2$ and $\bm \Psi = \bm U_1^\dagger \bm U_2$ that
\begin{align*}
 \norm{\widetilde{\bm{\Psi}}-\bm\Psi} 
& = \norm{{\bm U}_1^\dagger \breve{\bm U}_2 - \bm{U}_1^\dagger \bm{U}_2 } \\   
& = \norm{\bm {U}_1^\dagger (\breve{\bm U}_2 - \bm{U}_2)} \\ 
& \leq \norm{\bm {U}_1^\dagger} \cdot \norm{\breve{\bm U}_2 - \bm{U}_2} \\ 
& = \norm{\bm {U}_1^\dagger} \cdot \norm{(\bm{G}_2 -\bm{G}_1)  {\bm{\Phi}_2} \bm{P}} \\
& \overset{\mathrm{i})}{=} \norm{\bm {U}_1^\dagger} \cdot \norm{(\bm{G}_2 -\bm{G}_1)  {\bm{\Phi}_2}  \bm{\Phi}^{\dagger}\bm{G}^\dagger \bm U} \\
& \leq \norm{\bm {U}_1^\dagger} \cdot \norm{\bm{G}_2 -\bm{G}_1} \cdot \norm{{\bm{\Phi}_2} \bm{\Phi}^\dagger} \cdot \norm{\bm{G}^\dagger} \cdot \norm{\bm U},
\end{align*}
where the identity in i) holds since $\bm P$ is expressed as $\bm{P} = \bm{\Phi}^{\dagger}\bm{G}^\dagger \bm U$. 
Then the claim in \eqref{eq:smd_Psibreve} follows since $\norm{\bm{\Phi}_1 \bm{\Phi}^\dagger} \leq \norm{\bm{\Phi} \bm{\Phi}^\dagger} \leq 1$. 
\end{IEEEproof}

It remains to derive an upper bound on $\kappa(\bm P)$. 
Note that $\bm P$ can be written as $\bm P = \bm{\Phi}^\dagger \bm G^{-1} \bm {U}$. 
Therefore, we have
\begin{equation}
\label{eq:ub_kappa_P}
\kappa(\bm{P}) 
\leq \kappa(\bm{\Phi}^\dagger) \kappa(\bm G^{-1}) \kappa(\bm {U}) 
= \kappa(\bm{\Phi}) \kappa(\bm G)
= \frac{\kappa(\bm{\Phi}) G_{\max}}{G_{\min}}. 
\end{equation}
Finally, it has been shown \cite[Equation~(III.1)]{li2020super} that
\begin{equation}
\label{eq:md_conv}
\mathrm{md}(\mathcal{T}, \widehat{\mathcal{T}}) 
\leq \frac{1}{2\Gamma} \cdot \mathrm{md}(\Lambda(\breve{\bm \Psi}), \Lambda(\widehat{\bm \Psi})).
\end{equation}
Plugging in \eqref{eq:smd_Psihat}, \eqref{eq:smd_Psibreve}, \eqref{eq:ub_kappa_P}, and \eqref{eq:md_conv} into \eqref{eq:trieq_Psi} provides the assertion.

\subsection{Proof of Theorem \ref{thm:fully_sampled_case}}

We invoke Proposition~\ref{prop:gen_case_deterministic} to prove Theorem \ref{thm:fully_sampled_case}. 
To this end, we show that a sufficient condition for \eqref{eq:prop_cond} given by 
\begin{align}
\label{eq:prop_cond2_fs}
\min_{\bm R \in O_S} \norm{\widehat{\bm U} - \bm{U} \bm R}    \leq \sqrt{2} \mathrm{dist}(\bm{\widehat{U}},\bm U)< \frac{\sigma_S(\bm{U}_1)}{2}
\end{align}
is satisfied. 
Once Proposition~\ref{prop:gen_case_deterministic} is invoked, we simplify its consequence in \eqref{eq:det_err_bnd} to obtain the desired error bound in \eqref{eq:thm:errbnd}.

First we show that \eqref{eq:thmcond_M_fs} and \eqref{eq:thmcond_L_fs} imply \eqref{eq:prop_cond2_fs}. 
Recall that the columns of $\bm{U} \in \mathbb{C}^{|\Omega| \times S}$ (resp. $\widehat{\bm{U}} \in \mathbb{C}^{|{\Omega}| \times S}$) correspond to the $S$-dominant eigenvectors of $\bm{R}_{\bm{Y}} = \frac{1}{L} \bm{Y} \bm{Y}^\mathsf{H}$ (resp. $\bm{R}_{\widehat{\bm{Y}}} = \frac{1}{L} \widehat{\bm{Y}} \widehat{\bm{Y}}^\mathsf{H}$ where $\widehat{\bm{Y}} = \bm{Y} + \bm{Z}$).  
This subspace estimation step via principal component analysis requires that $L\geq S$, which is implied by the condition in \eqref{eq:thmcond_L_fs}. 
The following lemma derives an upper bound on $\mathrm{dist}(\bm{\widehat{U}},\bm U)$  in \eqref{eq:prop_cond2_fs} by utilizing the classical Davis-Kahan theorem \cite{davis1970rotation} and the concentration of Gaussian matrices \cite{wainwright2019high,vershynin2018high}. 

\begin{lemma}
\label{lem:est_subsp_fs}
Suppose that the hypothesis of Theorem~\ref{thm:fully_sampled_case} holds except that $\Omega$ is fixed to an arbitrary deterministic set. Let 
\begin{align}
\label{eq:nsr_ratio}
\nu := \frac{\sigma^2}{G_{\min}^2 \sigma_S^2(\bm{\Phi}) \lambda_S(\bm{R}_{\bm{X}})}.
\end{align} where $\bm{R}_{\bm{X}} = \frac{1}{L}\bm{XX}^{\mathsf{H}}$.
Then there exist absolute constants $C_1$, $C_2$ and $c$ for which the following statement holds: 
If 
\begin{equation}
\label{eq:cond:lem:est_subsp_fs}
L 
\geq 
C_1 |\Omega| \nu \left( \rho^2 \kappa^2(\bm{\Phi}) \kappa(\bm{R}_{\bm{X}}) \vee \nu \right),
\end{equation}
then with probability at least $1-3e^{-cM}$ we have
\begin{equation}
\label{eq:res:lem:est_subsp_fs}
\mathrm{dist}(\bm{\widehat{U}},\bm{U}) 
\leq \frac{C_2  \sqrt{|\Omega| \nu (\rho^2 \kappa^2(\bm{\Phi}) \kappa(\bm{R}_{\bm{X}}) \vee \nu)}}{\sqrt{L}}.    
\end{equation}
\end{lemma}

\begin{IEEEproof}
Since $\bm{U}$ (resp. $\hat{\bm{U}}$) spans the dominant invariant space of $\bm{R}_{\bm{Y}}$ (resp. $\bm{R}_{\bm{\hat{Y}}}$), the Davis-Kahan $\sin\theta$ theorem \cite{davis1970rotation} provides a perturbation bound as follow: If 
\begin{equation} \label{DK-cond_fs}
 \norm{\bm{R}_{\widehat{\bm{Y}}} - \bm{R}_{\bm{Y}} - \sigma^2 \bm{I}_{|\Omega|}} 
\leq \frac{\lambda_S(\bm{R}_{\bm{Y}})}{2},
\end{equation}
then
\begin{align} \label{eq:daviskahan_fs}
\mathrm{dist}(\bm{\widehat{U}},\bm{U}) 
\leq \frac{2\norm{\bm{R}_{\widehat{\bm{Y}}} - \bm{R}_{\bm{Y}} - \sigma^2 \bm{I}_{|\Omega|}}}{\lambda_S(\bm{R}_{\bm{Y}})}.
\end{align}
We show that with high probability, \eqref{eq:cond:lem:est_subsp_fs} implies \eqref{DK-cond_fs}, and \eqref{eq:daviskahan_fs} implies \eqref{eq:res:lem:est_subsp_fs}.
We start by using the triangle inequality and the definition of $\widehat{\bm{Y}}$, then the left-hand side of \eqref{DK-cond_fs} satisfies 
\begin{align*}
& L \norm{\bm{R}_{\widehat{\bm{Y}}} - \bm{R}_{\bm{Y}} - \sigma^2 \bm{I}_{|\Omega|}} 
 = \norm{\widehat{\bm{Y}} \widehat{\bm{Y}}^* - \bm{Y} \bm{Y}^* - L  \sigma^2 \bm{I}_{|\Omega|}} \\
& \quad \leq \norm{ \bm{YZ}^*} + \norm{\bm{ZY}^*} + \norm{ \bm{ZZ}^* - L \sigma^2 \bm{I}_{|\Omega|}} \\
& \quad = 2 \norm{ \bm{YZ}^*} + \norm{ \bm{ZZ}^* - L \sigma^2 \bm{I}_{|\Omega|}}. 
\end{align*}
Due to the standard Gaussian concentration \cite{wainwright2019high,vershynin2018high} (also see \cite[Equation~(67)]{yang2022nonasymptotic}), it holds with probability at least $1 - 2e^{- c_2 |\Omega|}$ that 
\begin{equation}
\label{eq:event2:lem:est_subsp_fs}
\norm{ \bm{ZZ}^* - L \sigma^2 \bm{I}_{|\Omega|}}
\leq 8 \sigma^2 \max(\sqrt{|\Omega|L},|\Omega|). 
\end{equation}
Next, since $\bm{YZ}^*$ can be written as $\bm{C}^{1/2} \bm{\Upsilon}$ where $\bm{C} = \sigma^2 \bm{YY}^*$ and $\bm{\Upsilon} \in \mathbb{R}^{|\Omega| \times |\Omega|}$ is a random matrix whose entries are i.i.d. $\mathrm{Normal}(0,1)$, it follows that 
\begin{align}
\norm{\bm{YZ}^*} 
= \norm{\sigma (\bm{YY}^*)^{1/2}\bm{\Upsilon}} 
\leq & \sigma \norm{(\bm{YY}^*)^{1/2}}  \norm{\bm{\Upsilon}} \nonumber \\ 
\leq & 3 \sigma \norm{\bm{Y}} \sqrt{|\Omega|}
\label{eq:event3:lem:est_subsp_fs}
\end{align}
with probability $1- e^{-c_2 |\Omega|}$ \cite[Equation~(68)]{yang2022nonasymptotic}. 
We proceed conditioned on the events in \eqref{eq:event2:lem:est_subsp_fs} and \eqref{eq:event3:lem:est_subsp_fs}, all of which hold with probability $1-3e^{-c|\Omega|}$. 
Next we show that under these events, the assumption in \eqref{eq:cond:lem:est_subsp_fs} implies the condition in \eqref{DK-cond_fs}. 

On one hand, by utilizing \eqref{eq:event2:lem:est_subsp_fs} and \eqref{eq:event3:lem:est_subsp_fs}, we obtain that
\begin{align}
& 2 \norm{ \bm{YZ}^*} + \norm{ \bm{ZZ}^*  - L  \sigma^2 \bm{I}_{|\Omega|}} \nonumber \\ 
& \leq 6 \sigma \norm{\bm{Y}} \sqrt{|\Omega|} + 8 \sigma^2 \max(\sqrt{|\Omega|L},|\Omega|) \nonumber \\
& \leq 6 \sigma \norm{\bm{G}} \norm{\bm{\Phi}} \norm{\bm{X}} \sqrt{|\Omega|} + 8 \sigma^2 \max(\sqrt{|\Omega|L},|\Omega|) \nonumber \\
& = 6 \sigma G_{\max} \norm{\bm{\Phi}}  \norm{\bm{X}} \sqrt{|\Omega|} + 8 \sigma^2 \max(\sqrt{|\Omega|L},|\Omega|),
\label{eq:DK-num_fs}
\end{align}
where the last inequality stems from $\norm{\bm{Y}} = \norm{\bm{G \Phi X}} \leq \norm{\bm{G}} \norm{\bm{ \Phi}} \norm{\bm{X}}$ and $\norm{\bm{G}}  = G_{\max} $.
On the other hand, the numerator in right-hand side of \eqref{DK-cond_fs} is lower-bounded by 
\begin{equation} 
\label{eq:DK_denom_fs}
\lambda_S(\bm{R}_{\bm{Y}}) 
\geq 
{G_{\min}^2 \sigma_S^2(\bm \Phi) \lambda_S(\bm{R}_{\bm{X}})}.
\end{equation}
From the results in \eqref{eq:DK-num_fs} and \eqref{eq:DK_denom_fs}, we obtain a sufficient condition of \eqref{DK-cond_fs} given by
\begin{align} 
& \frac{6 \sigma G_{\max} \norm{\bm{ \Phi}} \norm{\bm{X}} \sqrt{|\Omega|}  + 8 \sigma^2 \max(\sqrt{|\Omega|L},|\Omega|)}{L} \nonumber \\ 
& \leq \frac{G_{\min}^2 \sigma_S^2(\bm \Phi) \lambda_S(\bm{R}_{\bm{X}})}{2}. 
\label{DK_cond2_fs}
\end{align}
Furthermore, since $\norm{\bm{X}} = \sqrt{L \lambda_1(\bm{R_X})}$, we obtain a sufficient condition of \eqref{DK_cond2_fs} given by
\begin{align*}
& \frac{G_{\min}^2 \sigma_S^2(\bm \Phi)  \lambda_S(\bm{R}_{\bm{X}})}{2} 
\geq \max \Bigg( \frac{16 \sigma^2 \sqrt{|\Omega|}}{\sqrt{L}}, \\
& \qquad \frac{16 \sigma^2|\Omega|}{L}, \frac{12 \sigma G_{\max} \sigma_1(\bm \Phi) \sqrt{\lambda_1(\bm{R_X}) |\Omega|}  }{\sqrt{L}} \Bigg),
\end{align*}
which is equivalently rewritten as 
\begin{equation}
\label{eq:L_3_ineq_fs}
\begin{aligned}
L & \geq C_1 |\Omega| \cdot \max \Bigg( \frac{ \sigma^2 G_{\max}^2 \sigma_1^2(\bm \Phi)  \lambda_1(\bm{R_X})}{G_{\min}^4 \sigma_S^4(\bm \Phi)\lambda_S^2(\bm{R_X}) }, \\ 
& \quad \frac{ \sigma^4 }{G_{\min}^4 \sigma_S^4(\bm \Phi)\lambda_S^2(\bm{R_X})}, 
\frac{ \sigma^2 }{G_{\min}^2 \sigma_S^2(\bm \Phi)\lambda_S(\bm{R_X})}\Bigg).
\end{aligned}
\end{equation}
for some absolute constant $C_1 > 0$. 
Then, since $\kappa(\bm{R_X}) = \frac{\lambda_1(\bm{R_X})}{\lambda_S(\bm{R_X})}$, $\kappa(\bm{\Phi}) = \frac{\sigma_1(\bm \Phi)}{\sigma_S(\bm \Phi)}$, $\rho = \frac{G_{\max}}{G_{\min}}$, and using the definition of $\nu$ in \eqref{eq:nsr_ratio}, we obtain that \eqref{eq:L_3_ineq_fs} is equivalently rewritten as 
\begin{equation}
\label{eq:L_3_ineq_fs_alt}
L \geq C_1 |\Omega| \max \left( \nu \kappa(\bm{R_X}) \kappa^2(\bm{\Phi}) \rho^2 , \nu^2 , \nu  \right).
\end{equation}
Since $\rho \geq 1$, $\kappa(\bm{R_X}) \geq 1$ and $\kappa^2(\bm{\Phi}) \geq 1$, we have $\nu \kappa(\bm{R_X}) \kappa^2(\bm{\Phi}) \rho^2 \geq \nu$. Therefore, the expression in \eqref{eq:L_3_ineq_fs_alt} further simplifies to \eqref{eq:cond:lem:est_subsp_fs}. 

Finally, plugging in the results by \eqref{eq:DK-num_fs} and \eqref{eq:DK_denom_fs} into \eqref{eq:daviskahan_fs} yields
\begin{align}
& \mathrm{dist}(\bm{\widehat{U}},\bm{U}) \nonumber \\ 
& \leq \frac{2}{L} \left(\frac{6 \sigma G_{\max} \norm{\bm{\Phi}} \norm{\bm{X}}  \sqrt{M} + 8 \sigma^2 \max(\sqrt{|\Omega|L},|\Omega|)}{G_{\min}^2 \sigma_S^2(\bm \Phi) \lambda_S(\bm{R}_{\bm{X}})} \right) \nonumber \\ 
& \leq \frac{4 \sqrt{|\Omega|}}{\sqrt{L}} \max \Bigg(\frac{6 \sigma \rho \kappa(\bm{\Phi}) \sqrt{\kappa(\bm{R_X})}  }{\sqrt{\lambda_S(\bm{R}_{\bm{X}})} G_{\min} \sigma_S(\bm \Phi)}, \nonumber \\ 
& \qquad \frac{8 \sigma^2 }{G_{\min}^2 \sigma_S^2(\bm \Phi) \lambda_S(\bm{R}_{\bm{X}})}, \frac{8 \sigma^2 \sqrt{|\Omega|/L}}{G_{\min}^2 \sigma_S^2(\bm \Phi) \lambda_S(\bm{R}_{\bm{X}})}  \Bigg).
\label{eq:intermed:dk_fs}
\end{align}
Using the definition of $\nu$, the result in \eqref{eq:intermed:dk_fs} can be further simplified to
\begin{align}
& \mathrm{dist}(\bm{\widehat{U}},\bm{U})
\leq \frac{C_2  \sqrt{|\Omega| \nu}}{\sqrt{L}} \nonumber \\ 
& \quad \cdot \left( \rho \kappa(\bm{\Phi}) \sqrt{\kappa(\bm{R_X})} \vee \sqrt{\nu} \vee \sqrt{\frac{\nu |\Omega|}{L}}  \right)    
\label{lem_last_step_fs}
\end{align}
for some constant $C_2 >0 $. 
Due to \eqref{eq:cond:lem:est_subsp_fs}, we have that $\sqrt{\frac{\nu |\Omega|}{L}}$ in \eqref{lem_last_step_fs} is dominated by $\rho \kappa(\bm{\Phi}) \sqrt{\kappa(\bm{R_X})}$. 
Therefore, the upper bound in \eqref{eq:res:lem:est_subsp_fs} is obtained. 
\end{IEEEproof}

Next, to show that 
\eqref{eq:thmcond_M_fs} and \eqref{eq:thmcond_L_fs} invoke Lemma~\ref{lem:est_subsp_fs}, we derive bounds on $\kappa(\bm{\Phi})$, $\sigma_S(\bm{\Phi})$, and $\sigma_S(\bm{U}_1)$ in interpretable forms by using the following lemmas.

\begin{lemma}[{Corollary of \cite[Theorem~2.3]{moitra2015super}}]
\label{lemma_kappa_Phi_fs}
Suppose that the hypothesis of Theorem~\ref{thm:fully_sampled_case} holds. Then we have
\begin{align}
\label{eq:kappa_phi_approx_fs}
\kappa(\bm \Phi) \leq \sqrt{2}
\end{align}
and
\begin{equation}
\label{eq:lb_sigma_s_Phi}
\sigma_S(\bm{\Phi}) \geq \sqrt{\frac{M}{2}}. 
\end{equation}
\end{lemma}

\begin{IEEEproof}
By the construction of $\Omega = \Omega_1 \cup \Omega_2$, we have that $\bm{\Phi}$ is a Vandermonde matrix. 
Since ${M} \geq \frac{3}{\Delta} + 1 \geq \frac{1}{\Delta}+1$, it follows from \cite[Theorem~2.3]{moitra2015super} that 
\[
\kappa(\bm \Phi)  \leq \sqrt{\frac{M+1/\Delta-1}{M-1/\Delta-1}} \leq \sqrt{2},
\] 
where the latter inequality holds if and only if $M \geq \frac{3}{\Delta}+1$. 
Furthermore, it has been shown in the proof of \cite[Theorem~2.3]{moitra2015super} that 
\[
\sigma_S^2(\bm{\Phi}) 
\geq M - \frac{1}{\Delta} - 1 
\geq \frac{M}{2},
\]
where the second inequality holds if and only if $M \geq \frac{2}{\Delta}+2$. 
\end{IEEEproof}

\begin{lemma} 
\label{lem:sing_U_1}
Suppose that the hypothesis under the Proposition~\ref{prop:gen_case_deterministic} holds. Then we have
\begin{equation}
\label{eq:sing_U_1_approx_fs}
\min(\sigma_S^2(\bm{U}_1),\sigma_S^2(\bm{U}_2)) \geq  1 - \frac{\rho^2 S }{\sigma_S^2(\bm{\Phi})} 
\end{equation}
\end{lemma} 
\begin{IEEEproof}
The distinction from \cite[Lemma~3]{li2020super} is that $\bm{U}$ spans the column space of $\bm{G} \bm{\Phi}$ instead of $\bm{\Phi}$. 
Following the argument in the proof of \cite[Lemma~3]{li2020super}, we consider alternative expressions of $\bm{U}$. 
Let $\bm{w}_2^*$ denote the first row of $\bm{U}$ and $\bm{U}_2$ be the submatrix with the other $S-1$ rows. 
Let $\bm{w}_1^*$ and $\bm{U}_1$ be defined similarly with the last row. 
Then $\bm{U}$ can be equivalently rewritten as
\begin{align*}
\bm{U} 
= 
\begin{bmatrix}
\bm{w}_2^* \\ \bm{U}_2 
\end{bmatrix} 
= 
\begin{bmatrix}
\bm{U}_1 \\ \bm{w}_1^* \\
\end{bmatrix}.
\end{align*} 
For $j = 1,2$, since 
\[
\bm{w}_j \bm{w}_j^* + \bm{U}_j^\mathsf{H} \bm{U}_j
= 
\begin{bmatrix}
    \bm{w}_j & \bm{U}_j^* 
\end{bmatrix} \begin{bmatrix}  \bm{w}_j^* \\ \bm{U}_j \end{bmatrix}
= \bm{U}^\mathsf{H}\bm{U} = \bm{I}_S,
\]
it follows that
\[
\bm{U}_j^\mathsf{H} \bm{U}_j 
= \bm{I}_S - \bm{w}_j \bm{w}_j^*.
\]
Therefore, we have
\[
( \bm{U}_j^\mathsf{H} \bm{U}_j) \bm{w}_j = (\bm{I}_S - \bm{w}_j \bm{w}_j^*) \bm{w}_j = (1 - \norm{\bm{w}_j}^2_2 ) \bm{w}_j, 
\]
which implies that $\bm{w}_j$ is the eigenvector of $\bm{U}_j^\mathsf{H} \bm{U}_j$ associated with eigenvalue $1 - \norm{\bm{w}_j}^2_2$.
Furthermore, since $\bm{U}_j^\mathsf{H} \bm{U}_j$ is self-adjoint, the other eigenvectors are orthogonal to $\bm{w}_j$ and their corresponding eigenvalues are $1$. 
In other words, $\bm{U}_j$ satisfies 
\begin{align}
&\sigma_1({\bm{U}_j}) = \sigma_2({\bm{U}_j}) = \cdots 
= \sigma_{S-1}({\bm{U}_j}) = 1 
\quad \text{and} \quad \nonumber \\ 
& \sigma_S({\bm{U}_j}) = \sqrt{1 - \norm{\bm{w}_j}^2_2}. 
\label{eq:intermediat_sing_val_U_j_fs}
\end{align}

Let $\bm{e}_k$ denote the $k$th column of $\bm{I}_{|\Omega|}$ for $k \in [|\Omega|]$. Then by the variational characterization of the $\ell_2$ norm, we have 
\begin{align*}
\norm{\bm{w}_2}_2 = \sup_{\bm{z} \neq \bm{0}} \frac{\left| \bm{e}_1^\mathsf{T} \bm{U} \bm{z} \right|}{\norm{\bm{z}}_2}.
\end{align*}
Since $\bm{U}$ and $\bm{G}\bm{\Phi}$ share the same column space and they have the full column rank, for each $\bm{z} \neq \bm{0}$, there is a nonzero column vector $\bm{x}$ such that $\bm{U} \bm{z} = \bm{G} \bm{\Phi} \bm{x}$. 
Furthermore, since $\bm{U}^\mathsf{H} \bm{U} = \bm{I}_S$, we have $\norm{\bm{z}}_2 = \norm{\bm{U} \bm{z}}_2 = \norm{\bm{G} \bm{\Phi} \bm{x}}_2$. 
For any nonzero $\bm{x}$, we have
\begin{equation}
\label{eq:intermediate_ww}
\begin{aligned}
\frac{\left| \bm{e}_1^\mathsf{T} \bm{U} \bm{z} \right|}{\norm{\bm{z}}_2}
& = 
\frac{|\bm{e}_1^\mathsf{T} \bm{G \Phi x}|}{\norm{\bm{G \Phi x}}_{2}} \\
 & \leq 
\frac{\norm{\bm{G \Phi x}}_{\infty}}{\norm{\bm{G \Phi x}}_{2}} \\
& \overset{\mathrm{(a)}}{\leq} \frac{\norm{\bm{G} : \ell_\infty^{|\Omega|} \rightarrow \ell_\infty^{|\Omega|}}  \norm{\bm{\Phi} : \ell_1^S \rightarrow \ell_\infty^{|\Omega|}} \norm{\bm{x}}_1}{\sigma_S(\bm{G}) \sigma_S(\bm{\Phi}) \norm{\bm{x}}_2} \\ \nonumber 
&\overset{\mathrm{(b)}}{\leq} \frac{G_{\max} \sqrt{S}}{G_{\min} \sigma_S(\bm{\Phi})}
= \frac{\rho \sqrt{S}}{\sigma_S(\bm{\Phi})},    
\end{aligned}
\end{equation}
where (a) follows from the definition of the operator norm and the minimum singular value; and (b) holds since the diagonal matrix $\bm{G}$ satisfies $\norm{\bm{G} : \ell_\infty^{|\Omega|} \rightarrow \ell_\infty^{|\Omega|}} = G_{\max}$ and $\sigma_S(\bm{G}) = G_{\min}$, $\norm{\bm{\Phi} : \ell_1^S \rightarrow \ell_\infty^{|\Omega|}}$ denotes the largest magnitude of $\bm{\Phi}$, which is $1$, and $\norm{\bm{x}}_1 \leq \sqrt{S} \norm{\bm{x}}_2$ for all $\bm{x} \in \mathbb{C}^S$. 
By taking the supremum of \eqref{eq:intermediate_ww} over $\bm{x} \neq \bm{0}$, we obtain 
\begin{equation}
\label{eq:sing_Uj}
\norm{\bm{w}_2}_2
\leq \frac{\rho \sqrt{S}}{\sigma_S(\bm{\Phi})}.
\end{equation}
Finally, plugging in the result from \eqref{eq:sing_Uj} in \eqref{eq:intermediat_sing_val_U_j_fs} gives us the desired assertion. 
\end{IEEEproof}

The result of Lemma~\ref{lemma_kappa_Phi_fs} in \eqref{eq:lb_sigma_s_Phi} combined with the definition of $\nu$ in \eqref{eq:nsr_ratio} and the fact $|\Omega| = M+1$ due to the construction of $\Omega_1$ and $\Omega_2$ provides an upper bound on $|\Omega|\nu$ given by
\begin{equation}
\label{eq:M_nu_interim}
     |\Omega| \nu \leq \frac{2 \sigma^2}{G_{\min}^2 \lambda_S(\bm{R}_{\bm{X}})}.  
\end{equation}

We verify that \eqref{eq:thmcond_L_fs} combined with the inequalities in \eqref{eq:kappa_phi_approx_fs} and \eqref{eq:M_nu_interim} imply \eqref{eq:cond:lem:est_subsp_fs} in the form of 
\[
L \geq  \frac{C_1 \sigma^2}{G^2_{\min} \lambda_S(\bm{R}_{\bm{X}})} \cdot \left( \rho^2 \kappa(\bm{R}_{\bm{X}}) \vee \frac{ \sigma^2}{M G^2_{\min} \lambda_S(\bm{R}_{\bm{X}})} \right).
\]
Hence Lemma~\ref{lem:est_subsp_fs} is invoked. 
Then plugging in \eqref{eq:kappa_phi_approx_fs} and \eqref{eq:M_nu_interim} into the result of Lemma~\ref{lem:est_subsp_fs} in \eqref{eq:res:lem:est_subsp_fs} yields 
\begin{equation}
\label{eq:simplified_dk_fs}
\begin{aligned}
\mathrm{dist}(\bm{\widehat{U}},\bm{U}) 
& \leq \frac{C_2 \sigma}{G_{\min} \lambda_S^{1/2}(\bm{R}_{\bm{X}}) \sqrt{L} } \nonumber \\ 
& \cdot \left(  \rho \kappa^{1/2}(\bm{R}_{\bm{X}}) \vee \frac{\sigma}{G_{\min} \lambda_S^{1/2}(\bm{R}_{\bm{X}}) \sqrt{M} } \right)
\end{aligned}
\end{equation}
Therefore, we obtain a sufficient condition for \eqref{eq:prop_cond2_fs} from \eqref{eq:simplified_dk_fs} given by 
\begin{align*}
     \frac{C_3  \sqrt{|\Omega| \nu (\rho^2 \kappa^2(\bm{\Phi}) \kappa(\bm{R}_{\bm{X}}) \vee \nu)}}{\sqrt{L}} \leq \frac{\sigma_S(\bm{U}_1)}{2},
\end{align*}
which is equivalently rewritten as 
\begin{align}
\label{eq:L:interim_fs}
   L \geq  \frac{C_4 \sigma^2}{G^2_{\min} \lambda_S(\bm{R}_{\bm{X}})} \left( \frac{\rho^2 \kappa(\bm{R}_{\bm{X}})}{\left(1-\frac{2 \rho^2 S}{M} \right)} \vee \frac{ \sigma^2}{M G^2_{\min} \lambda_S(\bm{R}_{\bm{X}})}  \right).
\end{align}
This verifies that \eqref{eq:thmcond_L_fs} implies \eqref{eq:L:interim_fs} when \eqref{eq:kappa_phi_approx_fs} and \eqref{eq:M_nu_interim} are plugged. 
Therefore, we have shown that the condition in \eqref{eq:prop_cond2_fs} to invoke Proposition~\ref{prop:gen_case_deterministic} is satisfied. 
Finally, by utilizing the results in \eqref{eq:sing_U_1_approx_fs} and \eqref{eq:simplified_dk_fs}, the error bound by Proposition~\ref{prop:gen_case_deterministic} simplifies to \eqref{eq:thm:errbnd_fs}. 
This concludes the proof.

\subsection{Proof of Theorem~\ref{thm:main}}

We begin the proof with applying Proposition~\ref{prop:gen_case_deterministic} to the ESPRIT with random sub-arrays. 
Conditioned on the random index sets $\Omega_1$ and $\Omega_2$, Lemma~\ref{lem:est_subsp_fs} still applies.
However, the cardinality of $\Omega = \Omega_1 \cup \Omega_2$ is not a fixed number but a random variable. 
We will proceed under an event where $|\Omega|$ is upper bounded by a fixed number. 
Recall that $|\Omega| = \sum_{i=1}^{\widetilde{M}} \beta_i$ satisfies $\mathbb{E}[|\Omega|] = M$. 
Since $\beta_i$s are i.i.d. Bernoulli, by using Chernoff's bound (e.g. \cite[Exercise 2.3.5]{vershynin2018high}), we obtain that 
\begin{equation}
\label{eq:boundon|Omega|}
\left|
|\Omega| - M 
\right| \leq \frac{M}{10}
\end{equation}
holds with probability $1 - 2 e^{-c_1 M}$. The remainder of the proof assumes that \eqref{eq:boundon|Omega|} holds. 

As shown in the proof of Theorem~\ref{thm:fully_sampled_case}, the error bound in \eqref{eq:det_err_bnd} holds if \eqref{eq:prop_cond2_fs} is satisfied. We first shows that 
\eqref{eq:thmcond_M_tilde}, \eqref{eq:thmcond_M}, and \eqref{eq:thmcond_L} imply \eqref{eq:prop_cond2_fs}. 
Then the expression in \eqref{eq:det_err_bnd} will be simplified via the consequences of \eqref{eq:thmcond_M_tilde}, \eqref{eq:thmcond_M}, and \eqref{eq:thmcond_L}. 
Toward these goals, we use the following lemma to get tail bounds on $\kappa(\bm{\Phi})$, $\sigma_S(\bm{\Phi})$, and $\sigma_S(\bm{U}_1)$ when $\Omega_1$ and $\Omega_2$ are given as in \eqref{eq:Omega1_random_doublets}. 

\begin{lemma}
\label{lemma_kappa_Phi}
Let $\epsilon, \eta \in (0,1)$. 
Suppose that $\widetilde{M} > 1 + 1/\Delta$.
There exists an absolute constant $C > 0$ such that if 
\begin{equation}
\label{eq:sample_M}
M \geq C \epsilon^{-2} S \ln(S/\eta),
\end{equation}
then it holds with probability at least $1-\eta$ that
\begin{equation}
\label{eq:ub_kappa_Phi}
\kappa(\bm \Phi) 
\leq \sqrt{ \frac{2 \left( 1 + \epsilon + \left(\frac{1}{\Delta} - 1 \right) \frac{1}{\widetilde{M}} \right)}{ 1 - \epsilon - \left(\frac{1}{\Delta} + 1\right) \frac{1}{\widetilde{M}} } },
\end{equation}
\begin{equation}
\label{eq:lb_sigma_S_Phi}
\sigma_S(\bm \Phi) \geq   \sqrt{\left(1 - \epsilon - \left(\frac{1}{\Delta} + 1\right) \frac{1}{\widetilde{M}} \right )M}, 
\end{equation}
and
\begin{equation} 
\label{eq:lb_smin_U1}
\sigma_S(\bm U_1) \geq \frac{\widetilde{G}_{\min} \sqrt{1-\epsilon - \left(1 + \frac{1}{\Delta} \right) \frac{1}{\widetilde{M}} }}{\sqrt{2} \widetilde{G}_{\max} \sqrt{1+\epsilon + \left(\frac{1}{\Delta} - 1 \right) \frac{1}{\widetilde{M}} }}.
\end{equation}
\end{lemma}

\begin{IEEEproof}
Let $\bm{\Phi}_j = \bm{\Pi}_j \bm{\Phi}$ for $j = 1,2$. Then we have
\begin{equation}
\label{eq:decomp_gram_Phi_Phi1n2}
\bm \Phi^\mathsf{H} \bm \Phi 
= \bm \Phi_1^\mathsf{H} \bm \Phi_1 + \bm \Phi_2^\mathsf{H} \bm \Phi_2. 
\end{equation}
Below we will show that
\begin{equation}
\label{eq:concent_Phi}
\mathbb{P} \left(
\norm{
\bm{\Phi}_j^\mathsf{H} \bm{\Phi}_j 
- 
\mathbb{E} 
\bm{\Phi}_j^\mathsf{H} \bm{\Phi}_j
} 
> 
\epsilon M \right) 
\leq \frac{\eta}{2}, \quad j = 1,2.
\end{equation}
Moreover, for $j=1,2$, we define $\widetilde{\bm \Phi}_j \in \mathbb{C}^{\widetilde{M} \times S}$ by
\[
(\widetilde{\bm \Phi}_j)_{i,k} := e^{-\mathsf{j} \pi (2i-2+j) \tau_k / T}, \quad i \in [\widetilde{M}], ~ k \in [S].
\]
Then the expectation of $\bm{\Phi}_j^\mathsf{H} \bm{\Phi}_j$ reduces to 
\[
\mathbb{E} \bm{\Phi}_j^\mathsf{H} \bm{\Phi}_j = \frac{M}{\widetilde{M}} \widetilde{\bm \Phi}_j^\mathsf{H} \widetilde{\bm \Phi}_j.
\]
It has been shown \cite{moitra2015super} that the extreme singular values of the Vandermonde matrix $\widetilde{\bm \Phi}_j$ are bounded by
\begin{equation}
\label{eq:bnds_exp}
\begin{aligned}
\sqrt{\widetilde{M} - 1/\Delta - 1} 
& \leq \sigma_S(\widetilde{\bm \Phi}_j) \\
& \leq \sigma_1(\widetilde{\bm \Phi}_j) 
\leq \sqrt{\widetilde{M} + 1/\Delta - 1}.  
\end{aligned}
\end{equation}
Then \eqref{eq:concent_Phi} and \eqref{eq:bnds_exp} imply \eqref{eq:ub_kappa_Phi}. 

In the remainder of the proof, we show that \eqref{eq:concent_Phi} holds. 
Fix $j \in \{1,2\}$. 
Let $\bm{a}_i$ denote the $i$th column of $\widetilde{\bm \Phi}_j^\mathsf{H} \in \mathbb{C}^{K \times \widetilde{M}}$ for $i \in [\widetilde{M}]$. 
Then by the construction of $\Omega_j$, $\bm{\Phi}_j^\mathsf{H} \bm{\Phi}_j$ is written as
\[
\bm{\Phi}_j^\mathsf{H} \bm{\Phi}_j 
= \sum_{i=1}^{\widetilde{M}} \beta_i \bm{a}_i \bm{a}_i^\mathsf{H}.
\]
Therefore it suffices to show that 
\begin{equation}
\label{eq:concent_PhiTPhi1}
\mathbb{P} 
\left( 
\left\|
\sum_{i=1}^{\widetilde{M}} \left(\beta_i - \frac{M}{\widetilde{M}} \right) \bm{a}_i \bm{a}_i^\mathsf{H}
\right\| > \epsilon M
\right) \leq \frac{\eta}{2}.
\end{equation}
For brevity, we use the following shorthand notation: 
\[
\bm \Xi_i = \left( \beta_i - \frac{M}{\widetilde{M}} \right) 
\bm{a}_i \bm{a}_i^\mathsf{H}, \quad i \in [\widetilde{M}]. 
\]
Then $\mathbb{E} \bm \Xi_i = \bm 0$ and the spectral norm of $\bm \Xi_i$ is upper-bounded as
\begin{align*}
\norm{\bm \Xi_i} 
\leq \norm{\bm{a}_i}_2^2 + \frac{M}{\widetilde{M}} \norm{\bm{a}_i}_2^2 
\leq 2 S. 
\end{align*}
Furthermore, since
\begin{align*}
\mathbb{E} [\bm \Xi_i^\mathsf{H} \bm \Xi_i] 
& = \mathbb{E}\left[ \left( \beta_i + \frac{M^2}{\widetilde{M}^2} - \frac{2 M \beta_i}{\widetilde{M}} \right) \norm{\bm{a}_i}_2^2 \bm{a}_i \bm{a}_i^\mathsf{H} \right] \\
& = \frac{M S}{\widetilde{M}} \left( 1 - \frac{M}{\widetilde{M}} \right) \bm{a}_i \bm{a}_i^\mathsf{H},
\end{align*}
it follows that
\begin{align*} 
\left\| \sum_{i=1}^{\widetilde{M}} \mathbb{E} [\bm \Xi_i^\mathsf{H} \bm \Xi_i] \right\| & \leq \frac{M S}{\widetilde{M}} \norm{\widetilde{\bm{\Phi}}_j^\mathsf{H} \widetilde{\bm{\Phi}}_j}.
\end{align*}
Therefore, due to the matrix Bernstein inequality \cite{tropp2012user}, it holds with probability at least $1-\eta/2$ that
\begin{equation}
\label{eq:ub_Xi_dev}
\begin{aligned}
& \left \| \sum_{i=1}^{\widetilde{M}} \bm \Xi_i \right \| \\
& \leq \max \Bigg( \frac{2 \norm{\widetilde{\bm{\Phi}}_j} \sqrt{M S \ln(2S/\eta)}}{\sqrt{\widetilde{M}}}, \frac{4 S \ln(2S/\eta)}{3} \Bigg) \nonumber \\
& \overset{\mathrm{i})}{\leq} M \max \Bigg( \frac{4 S \ln(2S/\eta)}{3 M}, \\
& \quad \frac{2 \sqrt{S (1 + 1/(\Delta \widetilde{M}) - 1/\widetilde{M}) \ln(2S/\eta)}}{\sqrt{M}} \Bigg) \nonumber \\
& \overset{\mathrm{ii})}{\leq} \epsilon M, \end{aligned}
\end{equation}
where i) follows from the fact that $\norm{\widetilde{\bm \Phi}_j} \leq \sqrt{\widetilde{M} + 1/\Delta - 1}$ and ii) holds since $\widetilde{M} > 1/\Delta$ as one selects the constant $C$ in \eqref{eq:sample_M} sufficiently large. 

By Weyl's inequality, we have
\begin{align}
\label{eq:ub_lambda1_Phi_j}
\lambda_1(\bm{\Phi}_j^\mathsf{H} \bm{\Phi}_j) 
& \leq \lambda_1(\mathbb{E} \bm{\Phi}_j^\mathsf{H} \bm{\Phi}_j)
+ \norm{\bm{\Phi}_j^\mathsf{H} \bm{\Phi}_j - \mathbb{E} \bm{\Phi}_j^\mathsf{H} \bm{\Phi}_j}\\ \nonumber
& \leq \left( 1 + \frac{1}{\Delta \widetilde{M}} - \frac{1}{\widetilde{M}} + \epsilon \right) M, 
\end{align}
where the second inequality follows from \eqref{eq:bnds_exp} and \eqref{eq:ub_Xi_dev}. 
Similarly, we also have 
\begin{align}
\label{eq:lb_lambdaS_Phi_j}
\lambda_S(\bm{\Phi}_j^\mathsf{H} \bm{\Phi}_j) 
& \geq \lambda_S(\mathbb{E} \bm{\Phi}_j^\mathsf{H} \bm{\Phi}_j)
- \norm{\bm{\Phi}_j^\mathsf{H} \bm{\Phi}_j - \mathbb{E} \bm{\Phi}_j^\mathsf{H} \bm{\Phi}_j}\\ \nonumber
& \geq \left( 1 - \frac{1}{\Delta \widetilde{M}} - \frac{1}{\widetilde{M}} - \epsilon \right) M. 
\end{align}
Therefore, by \eqref{eq:decomp_gram_Phi_Phi1n2}, we obtain 
\begin{align*}
\lambda_1(\bm{\Phi}^\mathsf{H} \bm{\Phi}) 
& \leq 2 \left( 1 + \frac{1}{\Delta \widetilde{M}} - \frac{1}{\widetilde{M}} + \epsilon \right) M
\end{align*}
and 
\begin{align*}
\lambda_S(\bm{\Phi}^\mathsf{H} \bm{\Phi}) 
& \geq \left( 1 - \frac{1}{\Delta \widetilde{M}} - \frac{1}{\widetilde{M}} - \epsilon \right) M. 
\end{align*}
The assertions in \eqref{eq:ub_kappa_Phi} and \eqref{eq:lb_sigma_S_Phi} follow from the above results. 

We proceed to the derivation of \eqref{eq:lb_smin_U1}. 
Note that $\bm U$ can be written as 
\[
\bm U = \bm G \bm \Phi (\bm \Phi^\mathsf{H} \bm G^\mathsf{H} \bm G \bm \Phi)^{-1/2} \bm R
\]
for a rotation matrix $\bm R \in \mathbb{C}^{S \times S}$. 
Let $\bm{G}_j = \bm{\Pi}_j \bm{G} \bm{\Pi}_j^\top$ for $j = 1,2$. Then we have
\[
\bm \Phi^\mathsf{H} \bm G^\mathsf{H} \bm G \bm \Phi 
= \bm \Phi_1^\mathsf{H} \bm G_1^\mathsf{H} \bm G_1 \bm \Phi_1 + \bm \Phi_2^\mathsf{H} \bm G_2^\mathsf{H} \bm G_2 \bm \Phi_2. 
\]
Therefore, the minimum singular value of $\bm U_1 = \bm G_1 \bm \Phi_1$ satisfies
\begin{align}
\sigma_S(\bm{U}_1) 
& \geq 
\frac{\sigma_S(\bm{G}_1 \bm{\Phi}_1)}{\sqrt{2} \max\{  \norm{\bm{G}_1 \bm{\Phi}_1}, \norm{\bm{G}_2 \bm{\Phi}_2} \}} \nonumber \\
& \geq 
\frac{\sigma_S(\bm{G}_1) \sigma_s(\bm{\Phi}_1)}{\sqrt{2} \max\{ \norm{\bm{G}_1}, \norm{\bm{G}_2} \} \cdot \max\{ \norm{\bm{\Phi}_1}, \norm{\bm{\Phi}_2} \}} \nonumber \\
& \geq 
\frac{\widetilde{G}_{\min} \sigma_S(\bm{\Phi}_1)}{\sqrt{2} \widetilde{G}_{\max} \norm{\bm{\Phi}_1}} \label{eq:lb_smin_U1_step}
\end{align}
where the last inequality follows from the fact that $\norm{\bm{\Phi}_1} = \norm{\bm{\Phi}_2}$, which is due to the construction in \eqref{eq:rotinv_Phi}, $\sigma_s(\bm G_1) \geq G_{\min} \geq \widetilde{G}_{\min}$, and $\max\{ \norm{\bm{G}_1}, \norm{\bm{G}_2} \} \leq G_{\max} \leq \widetilde{G}_{\max}$. 
Plugging in \eqref{eq:ub_lambda1_Phi_j} and \eqref{eq:lb_lambdaS_Phi_j} into \eqref{eq:lb_smin_U1_step} yields the last assertion in \eqref{eq:lb_smin_U1}.
\end{IEEEproof}

Let the parameters in Lemma~\ref{lemma_kappa_Phi} be set to $\epsilon=0.1$ and $\eta = 1/M$. 
By choosing $C_1$ in \eqref{eq:thmcond_M} sufficiently large, the condition in \eqref{eq:sample_M} is satisfied and hence Lemma~\ref{lemma_kappa_Phi} is invoked. 
Furthermore, since \eqref{eq:thmcond_M_tilde} implies $\widetilde{M} > \frac{3}{\Delta}+3$, the results of Lemma~\ref{lemma_kappa_Phi} yield
\begin{equation}
   \label{eq:kappa_phi_approx}
  \kappa(\bm \Phi) \leq 4,   
\end{equation} 
\begin{equation}
\label{eq:approx_sigma_s_phi_rd}
  \sigma_S(\bm \Phi) \geq \sqrt{\frac{M}{5}},
\end{equation} and 
\begin{equation}
\label{eq:sing_U_1_approx_rd}
  \sigma_S(\bm U_1) \geq \frac{\widetilde{G}_{\min} }{\sqrt{13} \widetilde{G}_{\max}} = \frac{1}{\sqrt{13} \widetilde{\rho}}
\end{equation}
hold with probability at least $1-M^{-1}$.
We proceed with the remainder of the proof conditioned on the event that  \eqref{eq:kappa_phi_approx}, \eqref{eq:approx_sigma_s_phi_rd} and \eqref{eq:sing_U_1_approx_rd} hold. 

The results in \eqref{eq:boundon|Omega|} and \eqref{eq:approx_sigma_s_phi_rd} combined with the definition of $\nu$ in \eqref{eq:nsr_ratio} provide an upper bound on $|\Omega| \nu$ given by 
\begin{equation}
\label{eq:boundon_Omega_nu_rd}
     |\Omega| \nu \leq \frac{C' \sigma^2}{  G_{\min}^2 \lambda_S(\bm{R}_{\bm{X}})}.
\end{equation}

We next verify that the condition in \eqref{eq:cond:lem:est_subsp_fs} holds under the inequalities in \eqref{eq:kappa_phi_approx} and  \eqref{eq:boundon_Omega_nu_rd} combined with \eqref{eq:thmcond_L}. 

Therefore Lemma~\ref{lem:est_subsp_fs} is invoked.  
Furthermore, plugging in \eqref{eq:kappa_phi_approx} and \eqref{eq:boundon_Omega_nu_rd} into the result \eqref{eq:res:lem:est_subsp_fs} of Lemma~\ref{lem:est_subsp_fs} yields 
\begin{equation}
\label{eq:simplified_dk_rd}
\begin{aligned}
\mathrm{dist}(\bm{\widehat{U}},\bm{U}) 
& \lesssim \frac{\sigma}{\sqrt{L}\widetilde{G}_{\min} \lambda_S^{1/2}(\bm{R}_{\bm{X}})} \\ 
& \cdot \left(  \widetilde{\rho} \kappa^{1/2}(\bm{R}_{\bm{X}}) \vee \frac{\sigma}{\widetilde{G}_{\min} \lambda_S^{1/2}(\bm{R}_{\bm{X}}) \sqrt{M } } \right).
\end{aligned}
\end{equation}
Additionally, we also verify that combining \eqref{eq:sing_U_1_approx_rd} and \eqref{eq:simplified_dk_rd} provides a sufficient condition for \eqref{eq:prop_cond2_fs} given in the form of
\[
L \geq \frac{C \sigma^2}{\widetilde{G}^2_{\min} \lambda_S(\bm{R}_{\bm{X}})} \left( \widetilde{\rho}^4 \kappa(\bm{R}_{\bm{X}}) \vee \frac{ \sigma^2 \widetilde{\rho}^2}{M \widetilde{G}^2_{\min} \lambda_S(\bm{R}_{\bm{X}})}  \right),
\]
which is implied by \eqref{eq:thmcond_L}.
Finally, the error bound by Proposition~\ref{prop:gen_case_deterministic} simplifies to that in Theorem~\ref{thm:main} by utilizing the results in \eqref{eq:kappa_phi_approx}, \eqref{eq:sing_U_1_approx_rd}, and \eqref{eq:simplified_dk_rd}. 
This concludes the proof.


\end{document}